\documentclass[aps,prd,superscriptaddress,floatfix,nofootinbib,preprintnumbers,eqsecnum]{revtex4-2}

\usepackage[utf8]{inputenc}
\usepackage[dvips]{graphicx}
\usepackage[dvipsnames]{xcolor}

\usepackage{color}
\usepackage{relsize}
\usepackage{graphics}
\usepackage{epstopdf}
\usepackage{hyperref}
\hypersetup
 {
   colorlinks,
   linkcolor={blue!80!black},
   citecolor={blue!70!black},
   urlcolor={blue!70!black}
 }
\usepackage{mathrsfs}
\usepackage{ragged2e}
\usepackage{amssymb}
\usepackage{subcaption}
\usepackage{array}
\usepackage{ stmaryrd}
\usepackage{subcaption}

\usepackage[normalem]{ ulem }
\usepackage{amsthm}
\usepackage{amsmath}
\usepackage{hyperref}
\usepackage{cancel}
\usepackage{verbatim} 

\newcommand{\phik}{\overline{ \phi}_k}
\newcommand{\phivev}{\overline{\phi}}
\newcommand{\be}{\begin{equation}}

\newcommand{\ee}{\end{equation}}
\newcommand{\bea}{\begin{eqnarray}}
\newcommand{\eea}{\end{eqnarray}}

\newcommand{\ie}{{\em i.e.~}}

\newcommand{\dd}{\text{d}}

\newcommand{\beq}{\begin{equation}}
\newcommand{\eeq}{\end{equation}}
\newcommand{\beqn}{\begin{eqnarray}}
\newcommand{\eeqn}{\end{eqnarray}}
\newcommand{\nn}{\nonumber}

\newcommand{\intdx}[2] 
{\left[ #1 \right]_{#2}} 
\newcommand{\ddelta}[2]{\frac{\delta #1}{\delta #2}}

\newcommand{\calG}{{\mathcal G}}
\newcommand{\calO}{{\mathcal O}}

\def\calG{{\cal G}}

\usepackage[dvips]{graphicx}
\usepackage{xcolor}
\usepackage{color}
\usepackage{graphics}
\usepackage{epstopdf}
\usepackage{hyperref}
\usepackage{mathrsfs}
\usepackage{amssymb}
\usepackage{epstopdf}
\usepackage{mathrsfs}
\usepackage{amssymb}
\usepackage{verbatim}
\usepackage{url}
\usepackage{amsmath}
\usepackage{hyperref}
\usepackage{psfrag}
\usepackage{amsmath}
\usepackage{amsfonts}
\usepackage{amssymb}
\usepackage{xcolor}
\usepackage{graphicx}
\usepackage{color}
\usepackage{amsthm}
\usepackage{oldgerm}
\usepackage{cancel}

\usepackage{chngpage}
\usepackage{calc}
\usepackage{graphicx}
\usepackage{array}
\usepackage{tikz}
\usepackage{pgfplots}
\usetikzlibrary{trees}
\usetikzlibrary{decorations.pathmorphing}
\usetikzlibrary{decorations.markings}

\usetikzlibrary{decorations.text}
   \definecolor{greeen}{rgb}{0.03,0.54,0.23}
\definecolor{test}{rgb}{0.03,0.74,0.33}
\definecolor{viol}{rgb}{0.44,0,0.94}
\definecolor{or}{rgb}{0.9,0.6,0}
\tikzset{
    photon/.style={decorate, decoration={snake, amplitude=2pt}, draw=green},
    photon2/.style={decorate, decoration={snake, amplitude=2pt}, draw=viol},
    dark/.style={draw=greeen, postaction={decorate},
        decoration={markings}},
    dark0/.style={draw=greeen, postaction={decorate},
        decoration={markings,mark=at position .5 with {\arrow[draw=greeen]{}}}},
antidark/.style={draw=greeen, postaction={decorate},
        decoration={markings}},
electron/.style={draw=viol, postaction={decorate},
        decoration={markings,mark=at position .5 with {\arrow[draw=viol]{>}}}},
        antielectron/.style={draw=viol, postaction={decorate},
        decoration={markings,mark=at position .5 with {\arrow[draw=viol]{<}}}},
        neutrino/.style={draw=orange, postaction={decorate},
        decoration={markings,mark=at position .5 with {\arrow[draw=orange]{>}}}},
        antineutrino/.style={draw=orange, postaction={decorate},
        decoration={markings,mark=at position .5 with {\arrow[draw=orange]{<}}}},
gluon/.style={decorate, draw=or,
        decoration={coil,amplitude=4pt, segment length=4pt}},
  ZZ/.style={decorate, decoration={snake}, draw=yellow},    
  scalar1/.style={dashed, draw=cyan, postaction={decorate}, decoration={markings,mark=at position .5 with {\arrow[draw=cyan]{>}}}},  
  scalar2/.style={dashed, draw=greeen, postaction={decorate}, decoration={markings,mark=at position .5 with {\arrow[draw=greeen]{>}}}},  
  scalar0/.style={decorate, dashed, draw=greeen}, 
  antiscalar1/.style={dashed, draw=cyan, postaction={decorate}, decoration={markings,mark=at position .5 with {\arrow[draw=cyan]{<}}}},  
  antiscalar2/.style={dashed, draw=greeen, postaction={decorate}, decoration={markings,mark=at position .5 with {\arrow[draw=greeen]{<}}}},  
  scalar0/.style={decorate, dashed, draw=greeen},  
 pseudoscalar/.style={decorate, dashed, draw=purple}, 
 scalar1no/.style={dashed, draw=cyan, postaction={decorate}}, 
 electronno/.style={draw=viol, postaction={decorate}},
   }

\usetikzlibrary{decorations, decorations.markings, decorations.pathmorphing, arrows, graphs, shapes.geometric, snakes}
\usetikzlibrary{shadings}

\pgfplotsset{
    colorbar style/.code={},
    every colorbar/.append style={xlabel={ {\Large $\delta$ }},
        /pgf/number format/fixed,grid=major
    }
}
\usetikzlibrary{shapes.misc}

\tikzset{cross/.style={cross out, draw=black, minimum size=2*(#1-\pgflinewidth), inner sep=0pt, outer sep=0pt},
cross/.default={1pt}}

\begin{document}
  \sloppy  

\preprint{IPPP/23/68}
\preprint{KCL-PH-TH/2023-66}

\vspace*{1mm}

\title{\Large Exact Schwinger Proper Time Renormalisation }
\def\andname{\hspace*{-0.5em}} 

\author{Steven Abel}
\email[Email address: ]{s.a.abel@durham.ac.uk}
\affiliation{Institute for Particle Physics Phenomenology, Durham University, Durham, DH1 3LE, United Kingdom}
\author{Lucien Heurtier}
\email[Email address: ]{lucien.heurtier@kcl.ac.uk}
\affiliation{Institute for Particle Physics Phenomenology, Durham University, Durham, DH1 3LE, United Kingdom}
\affiliation{Theoretical Particle Physics and Cosmology, King’s College London,\\ Strand, London WC2R 2LS, United Kingdom}

\begin{abstract}
\noindent We derive an exact version of the Schwinger Proper Time Renormalisation Group flow equation from first principles from the complete path integral, without using any perturbative expansion. We study the advantages of this flow equation as compared to the canonical Exact Renormalisation Group flow equation, which uses a regulator in momentum space. We use our flow equation to recover the convexity of the effective scalar potential in the IR limit and apply it to the study of false-vacuum decay. 
\end{abstract}
\maketitle

\tableofcontents

\section{Introduction}

The purpose of this paper is to develop an improvement of the Exact Renormalization Group (ERG) approach to renormalization.  The central feature of the ERG is the introduction of an energy scale $k $ into amplitudes using a specially dedicated infra-red (IR) regulator. In the original and canonical formulation of this procedure \cite{WETTERICH1991529,WETTERICH199390,Dupuis:2020fhh,Berges:2000ew}, the regulator is a momentum-space cut-off, constructed so as to integrate out all modes with a momentum $p\gg k$, thereby letting modes with $p<k$ interact in an effective theory defined at that energy scale. Remarkably, the procedure allows one to obtain exact 
information about the scaling of the effective potential with respect to the cut-off $k$ without the use of any perturbative expansion. 

\begin{figure*}
    \centering
\includegraphics[width=0.3\linewidth]{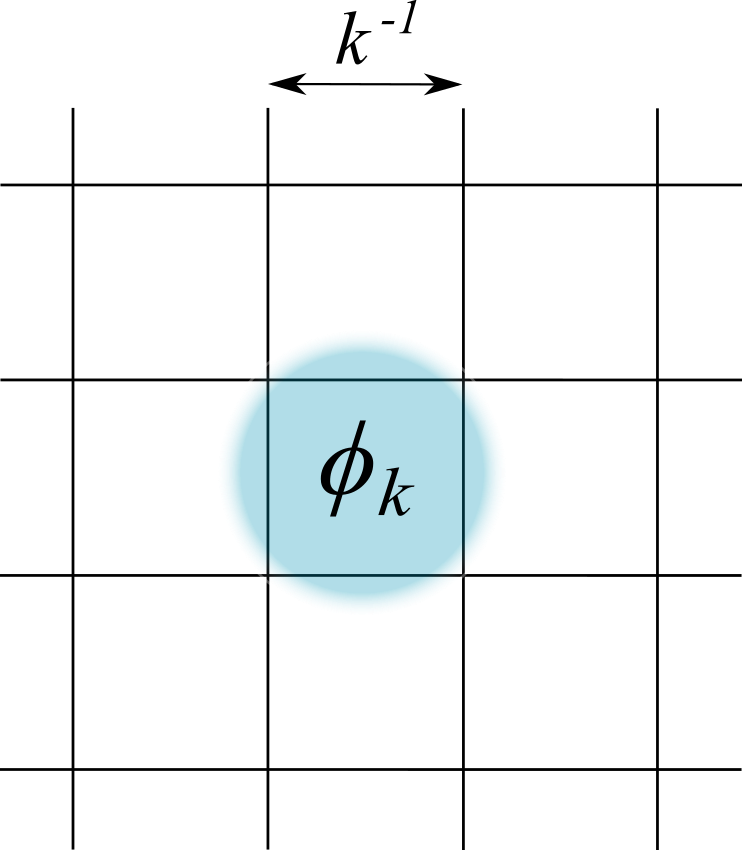}
    \caption{\label{fig:averaging_prelim} \footnotesize Averaging over cells to find the effective field theory.}
\end{figure*}

The geometric interpretation of the ERG is usually taken to be something resembling the picture shown in Fig.~\ref{fig:averaging_prelim}: since the effect of the cut-off is to yield an effective theory in which energetic modes have been integrated out, the theory at the scale $k$ is interpreted as consisting of fluctuations of coarse-grained fields built out of the cells that result from averaging the microscopic theory over a length-scale of $1/k$, as shown.
However, this broad-brush picture conflates momentum cut-off and inverse coarse-graining scale. Although the two are broadly equivalent, there is a distinction, which is the underlying motivation for the current paper: a four-momentum cut-off is not the same thing as a cut-off on propagation distance when the state is massive. In fact, to arrange a cut-off on the length scale, it is more consistent to use the proper time of a mode's propagation as the invariant measure of whether it belongs in the effective theory or not rather than its momentum-squared. Although such a Schwinger proper-time renormalisation group, typically referred to as PTRG,  has been advocated and discussed in several works \cite{Liao:1994fp,Liao:1995nm,Litim:2001up,Litim:2001hk,Zappala:2001nv,Litim:2002xm,Bonanno:2019ukb,Bonanno:2022edf}, the resulting flow equations do not have the all-orders exact status of the momentum cut-off prescription given in Refs~\cite{WETTERICH1991529,WETTERICH199390,Dupuis:2020fhh,Berges:2000ew}. The purpose of this paper is to derive such an exact equation for the renormalisation group flow using a proper-time regulator and also to demonstrate how this choice of regulator has significant utility when it comes to, for example, considering tunnelling and the related issue of the convexity of the potential. 

To flesh out the moral distinction between the two kinds of cut-off and give a little more context, let us begin by having a first look at the effect of a proper-time regulator. To this end, let us consider the Schwinger proper-time
formulation of the propagator, which is most often seen in the so-called ``world-line'' treatment of field theory \cite{Edwards:2019eby}. In this formulation, the (Euclideanized) scalar unregulated propagator can be written 
\begin{equation}
\label{eq:Schwing_prop}
\Delta(p)~ \, {\stackrel{\rm def}{=}} ~ \, \frac{1}{p^2+m^2} ~=~ \int_0^\infty \dd t \exp \left\{ -(p^2+m^2)t \right\} ~,
\end{equation}
where $t$ is the so-called proper time or world-line parameter. In perturbation theory, the geometric role of $t$ is literally to parameterise the proper-time 
propagation of the particle in the diagram along that particular propagator. Thus, a typical amplitude consists of a set of propagators glued together with vertices. To evaluate it, one is ultimately tasked with having to integrate over all the possible proper-times $t$ that appear in the diagram.

To introduce an infra-red (IR) cut-off on these proper times, which will quench contributions at large $t$ from any propagator, we insert a regulator $\calG^k (t)$ along the propagator's world-line as follows:
\beq 
\Delta(p)\quad\longrightarrow\quad\Delta^k(p) ~=~\int_0^\infty \dd t\ \calG^k(t)  \,
e^{-t(p^2+m^2)}\,.
\eeq
This {regularization procedure --- which, in spirit, resembles the regularization in momentum space introduced in Ref.~\cite{Polchinski:1983gv} --- } will, for example, lead to a quenching of the IR contribution of one-loop integrals, but we emphasize that this long-distance cut-off should be thought of as belonging to the propagators themselves rather than to any particular loop integral, and thus it affects the theory at all orders.  (It is worth remarking that  the regulator is defined to act on the Euclideanized propagators.)
As our proper-time cut-off  corresponds to quenching  the exponential in the propagator where $t\gg k^{-2}$, the  regulator $\calG^k (t)$ is required to have  the following properties:
\begin{eqnarray}\label{eq:Gprop}
\lim_{k^2 t\ll 1} \calG^k(t)&~=~&1~,\nonumber\\
\lim_{k^2 t\gg 1} \calG^k(  t)&~=~&0~.
\end{eqnarray}
The first condition ensures that one recovers the full theory in the limit $k\to 0$, whereas the second one removes the large-distance modes with $t\gg 1/k^2$. (Note that this cut-off is not to be confused with any ultra-violet (UV) short-distance cut-off at $t\to 0$ which may or may not be required for the UV finiteness of any particular diagram. As we shall see, the question of UV regularisation will not enter into our derivation.)

It is easy to see that, in contrast with the purely momentum-based cut-off, such a regulator not only integrates out of the theory short wavelength modes but also integrates out very massive states. For concreteness, consider the simplest example where $\calG^k$ is a Heaviside function $\calG^k(t) = \theta( k^{-2}-t )$. In this case, we find 
 \begin{equation} 
\Delta^k(p) ~=~\frac{1-e^{-(p^2+m^2)/k^2}}{p^2+m^2} ~.
 \end{equation}
Thus, we see that, as anticipated, only the short-wavelength/high-mass contributions can propagate in diagrams, with long wavelength light modes being quenched exponentially\footnote{It is worth mentioning that the subtraction implemented by the cut-off represents a non-locality that enters the theory at scales of order $1/k$. This non-locality is an unavoidable consequence of coarse-graining, given its interpretation as an averaging over cells of size $1/k$.
Such theories have been studied in their own right as non-local infinite-derivative field theories, albeit typically with the cut-off in that case implemented to regulate the UV~\cite{Padmanabhan:1996ap,Abel:2019ufz, Abel:2019zou}.}. 
From the perspective of space-time coordinates, the propagators with such a Heaviside cut-off are simply
\begin{align}
\Delta^k(x-y) ~&=~\int_0^{1/k^2}  \frac{dt}{(4\pi t)^2 }~ e^{-\frac{(x-y)^2}{4t} - m^2 t }~.
\end{align} 
For massless modes, setting $m=0$ gives an interaction in space-time coordinates that at short distances becomes the Coulomb interaction:
\begin{equation}
\Delta^k(x-y) ~=~ \frac{e^{-\frac{1}{4} k ^2 r^2}}{4 \pi^2 r^2}~,
\end{equation}
where $r=|x-y|$. For distances smaller than $1/k$ this is the regular Coulomb interaction, while the potential is exponentially suppressed for scales larger than $1/k$. Thus, for massless modes, the momentum/distance equivalence is exact. However, when the modes are massive, there is an interplay between the cut-off and the mass $m$. Indeed, when  $r<2m/k^2$, we can find the usual exponential decay expected of the propagator
(in the Euclidean region) from a saddle-point approximation:
\begin{equation}
\Delta^k(x-y) ~\approx ~
\frac{\sqrt{\frac{\pi }{2}} \sqrt{mr} e^{-m r}}{4 \pi ^2 r^{2}}~.
\end{equation}
On the other hand, when $r>2m/k^2$, there is no saddle point inside the domain of integration. In this case, we can neglect the $-m^2 t $ in the exponent and the state behaves essentially like the massless one, and so contributions at a scale $k$ with our regulated propagator are dominated by modes with wavelengths $r < {\rm Min}[1/m,1/k]$.

\begin{figure}
\centering
\includegraphics[keepaspectratio, width=0.49\textwidth]{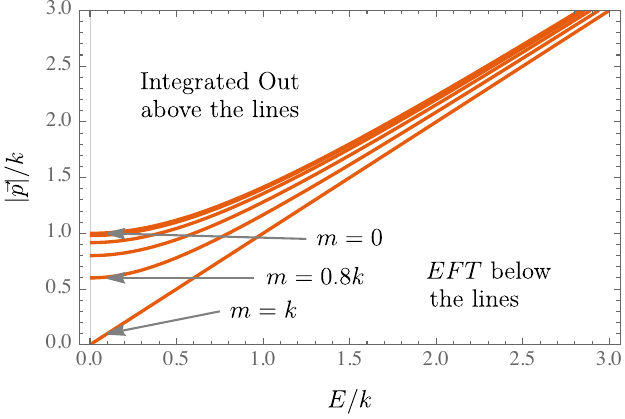}
\caption{The cut-off region according to the Schwinger proper-time regulator, with Minkowski signature.}
\label{fig:reg_scheme}
\end{figure}

To summarise then, we see that a proper-time cut-off differs from the pure momentum cut-off of Refs.~\cite{WETTERICH1991529,WETTERICH199390,Dupuis:2020fhh,Berges:2000ew} because it also excludes states with large masses from the effective theory regardless of their momentum. This distinction is shown in Fig.~\ref{fig:reg_scheme}, where we see that if we rotate back to Minkowski signature the cut-off regulated effective field theory (EFT) lives in the region below the $m$-dependent hyperbolae, with states that have  $|\vec p|^2 \geq E^2 + k^2 - m^2$ being  integrated out. Meanwhile, a purely momentum-dependent cut-off only integrates out the states above the $m=0$ line. From these general considerations, one can argue that from a coarse-graining perspective, the proper-time cut-off makes more physical sense: states that are much more massive than $k$ cannot propagate across a $1/k$ sized cell, so they have no business being in the effective theory. (A proper-time cut-off is also incidentally the only kind of coarse-graining that has a consistent analogue in conventional first quantized string theory \cite{Kiritsis:1994ta,Abel:2021tyt,Abel:2023hkk}.) We will have occasion to make further more detailed comparison throughout the paper. 

In the next section, we will use this regulator to derive an exact RG evolution equation for the PTRG. This equation is akin to that in Refs.~\cite{WETTERICH1991529,WETTERICH199390,Dupuis:2020fhh,Berges:2000ew}. We will use our equation to derive the flow in the so-called ``local potential approximation'' (LPA). 
Comparison with the standard momentum cut-off in Sec.~\ref{sec:comparison} reveals some advantages of the PTRG: in particular, we will see that, because the flow can be taken all the way to $k=0$, our equation can yield information about the onset of phase transitions and their associated tunnelling phenomena. In Sec.~\ref{sec:towards}, we prove that the evolution of the potential within the LPA as $k\to 0$ is always towards convexity regardless of the profile of the regulator function. Finally, in Sec.~\ref{sec:tunnelling} we look at tunnelling within this formalism. The aforementioned advantages allow us to see the potential "freeze" within the LPA without the need to stop the running by hand.

\section{Derivation of the exact RG}

To demonstrate the derivation of the exact renormalisation via the Schwinger cut-off method, let us consider real scalar theories:  the partition function of an interacting scalar theory is
\begin{equation}
Z[J]~=~N\int\mathcal D\phi\exp\left\{-
\intdx{
\frac{\partial\phi^2}{2}+\frac{m^2}{2}\phi^2 + V_I(\phi)}{}
+\intdx{ J\phi}{}
\right\}~,
\end{equation}
where we will throughout use $\intdx{.}{x} \equiv \int\dd^4 x (.)$ to stand for the integration over Euclidean space (dropping the subscript when there is no ambiguity), and where $N$ is a normalisation constant. In the absence of interactions, $V_I=0$, and we simply have the partition function of the free theory 
\begin{equation}
Z_0[J]~=~ N\int\mathcal D\phi\exp\left\{-\intdx{
\frac{\partial\phi^2}{2}+\frac{m^2}{2}\phi^2}{} +\intdx{J\phi}{} 
\right\}~.
\end{equation}
Now, clearly, a regulator on the propagators of the kind discussed in the introduction will enter into the entire theory through this free-field factor in the full $Z[J]$. In other words,  if we organise the theory into a free-field quadratic component with a partition function given by $Z_0$, which is acted upon by an interaction component to yield the full partition function, $Z$, then the regulator (and hence all the scale dependence) will arise from the free-field part. Therefore, our goal is to deduce the scaling (i.e.~the $k$-derivative) properties of $Z$ from the properties of the free-field component, $Z_0$, without ever having to explicitly compute anything in perturbation theory.

Our first step is thus to separate out this free-field component from $Z$ and introduce our regulator into it. Using standard manipulations, we first represent the interaction potential $V_I(\phi)$ as a function of operators in the usual manner,
\be
e^{-\intdx{V_I(\phi)}{}
}e^{\intdx{ J\phi}{} } ~ =~ e^{  -\intdx{ V_I\left(\frac{\delta}{\delta J}\right)}{} } e^{\intdx{ J\phi}{} }
\,.\ee
 This relation allows us to express the partition function of the interacting theory as an operator acting on  the free theory as follows:
\begin{eqnarray} 
Z[J] 
&~=~& e^{  -\intdx{ V_I\left(\frac{\delta}{\delta J}\right)}{} } Z_0[J]\,.
\label{eq:ZZZZ}
\end{eqnarray}
To express the free-field factor, we can
write it in terms of the Fourier-transformed currents,
\begin{eqnarray}
Z_0[\tilde J] 
&~=~& Z_0[0] \exp\left\{\frac{1}{2}\int\frac{d^4 p}{(2\pi)^4}\frac{\tilde J(p)\tilde J(-p)}{p^2+m^2}\right\}\,.
\end{eqnarray}
Together with  Eq.~\eqref{eq:ZZZZ}  this is basically the normal field-theoretic starting point. However, we now proceed to cut-off long-distance propagation in the theory as described in the Introduction, by regulating the free-field part, $Z_0$.
As previously, we first recast this factor  using the  Schwinger parametrization in Eq.~\eqref{eq:Schwing_prop}, giving
\begin{eqnarray}
Z_0[\tilde J] &~=~& Z_0[0] \exp\left\{\frac{1}{2}\int \frac{\dd^4 p}{(2\pi)^4} \int_0^\infty \dd t\ \tilde J(-p)\tilde J(p) e^{-t(p^2+m^2)}\right\}\,.
\end{eqnarray}
Following the discussion in the Introduction, we then truncate the infra-red (IR) modes of the theory by inserting the regulator $\mathcal G^k(t)$ into the Schwinger integral,
\be
Z_0^k[\tilde J] ~=~ Z_0[0] \exp\left\{\frac{1}{2}\int \frac{\dd^4 p}{(2\pi)^4} \int_0^\infty \dd t\ \mathcal G^k(t) \tilde J(-p)\tilde J(p) e^{-t(p^2+m^2)}\right\}~,
\label{eq:JJ}\ee
which yields a now regulated partition function: 
\be
Z^k[J]~\stackrel{\rm def}{=}~ e^{-\intdx{ V_I\left(\frac{\delta}{\delta J}\right)}{}  } Z_0^k[J]\,.
\ee

{This gives rise to  a $k$-dependent expectation value for the field. If we denote $W^k[J]=\log Z^k[J]$ 
we may write it as 
\be
\label{eq:phik}
\phik~\equiv~ \delta W^k[J]/\delta J \,.
\ee
Note that the $k$-suffix belongs with the expectation value not the field operator.
The regulated effective action is then defined in the same way as the effective action is normally defined, namely as a Legendre transform:
\be
\Gamma^k[\phik] ~\equiv~ -W^k[J] + \intdx{J\phik }{} \,
\label{eq:leg}
\ee
where $J$ is as usual identified as the source current for which the vacuum expectation value of $\phi$ satisfies the equation of motion.}

\begin{figure*}
    \centering
\includegraphics[width=0.5\linewidth]{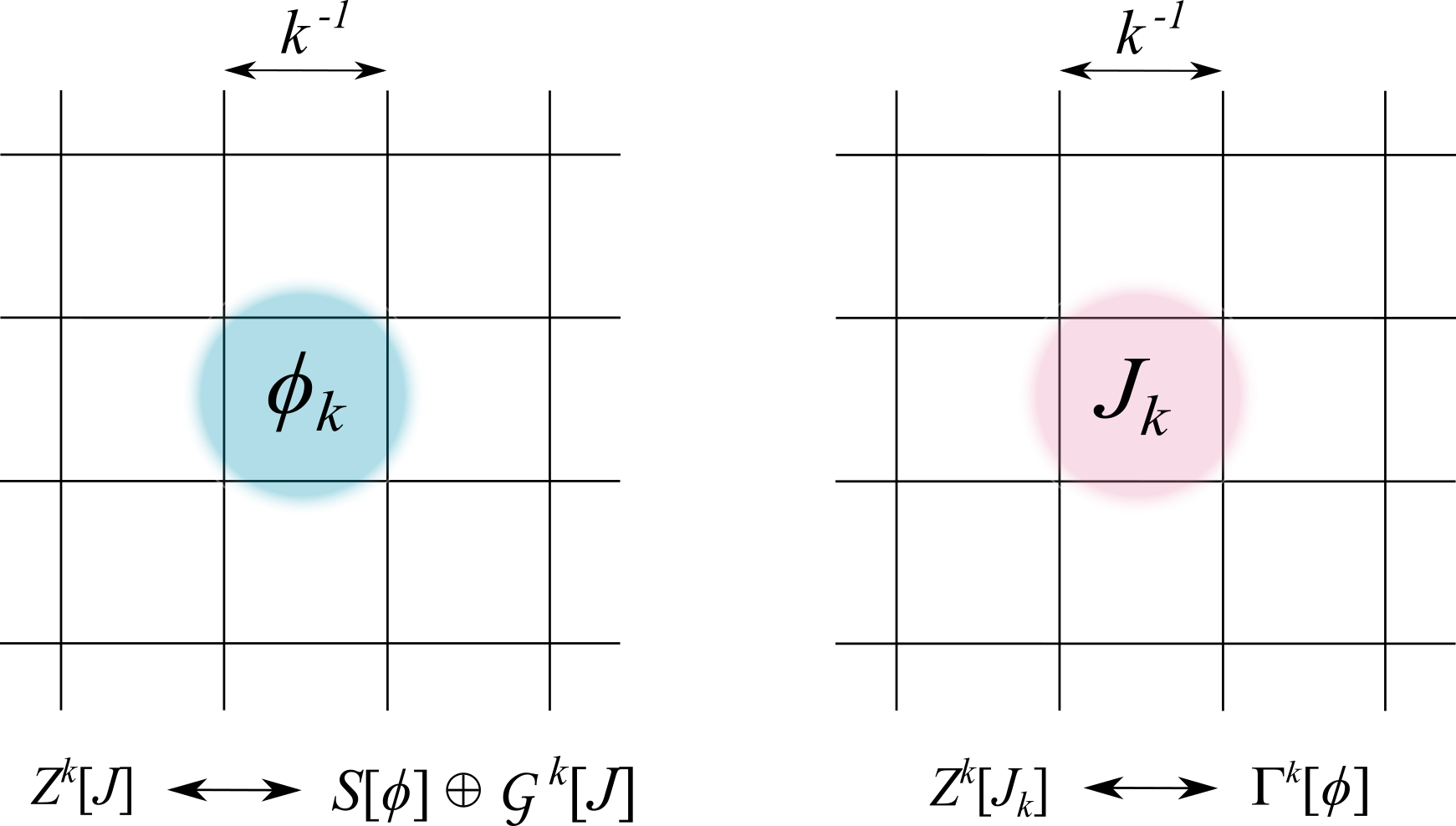}
    \caption{\label{fig:averaging} \footnotesize Equivalence between the theory described by the classical action $S[\phi]$ regulated at scale $k$ using $\mathcal G_k(t)$, in which the averaged field $\phi^k$ lives, and the effective field theory at scale $k$, in which the dynamics of $\phi$ is dictated by coarse-grained currents $J^k$, according to the effective action $\Gamma^k[\phi]$.}
\end{figure*}

It is worth pausing at this stage to reflect on the meaning of the regulated quantities. The current $J$ is, of course, just a source that we put into the path integral in order to get a response{ -- \ie it is the value of the current for which a particular field value solves the equations of motion: however the  field value corresponding to a fixed external source $J$ is now the $k$-dependent object, $\phik$, given by Eq.~\eqref{eq:phik}. 
It is more convenient for our purposes to instead ask what value of current would be needed to stimulate a certain fixed expectation value $\phivev$. 
This defines a $k$-dependent current that we will call $J_k$, which results from 
\be\label{eq:GammaDef}
\Gamma^k[\phivev] ~\equiv~ -W^k[J_k] + \intdx{J_k\phivev }{} \,
\ee
giving
\begin{align} 
\label{eq:derivs}
{J_k} &~\equiv~ \delta \Gamma^k[\phivev]/\delta \phivev ~.
\end{align}
Note that the physics is exactly the same, except that we are adjusting the current in order to maintain a particular value of $\phivev$
for the theory when it is being regulated at varying scales $k$.  For example, it is still the case that the vacuum of the theory at energy scale $k$ can be found by solving the $k$-dependent condition $ J^k =0$ which is simply a field equation for $\phivev$.}
As described in the Introduction, {these $k$-dependent quantities} correspond to expectation values, not of the original microscopic theory but of the theory after it has been broken into averaged $1/k$ sized cells. This is the practical implementation of the physical coarse-graining interpretation of Fig.~\ref{fig:averaging_prelim}, giving us the more detailed picture in Fig.~\ref{fig:averaging}.

{Note that, thanks to the properties of the regulator exhibited in Eq.~\eqref{eq:Gprop}, one recovers from this definition the full partition function and effective action, $Z^k[J]\to Z[J]$ and $\Gamma^k[\phivev]\to\Gamma[\phivev]$, in the limit $k\to 0$. This then leads to an implicit formula for the running of $\Gamma^k$, which is
\bea\label{eq:flow}
\partial_k \Gamma^k[\phivev ] ~&=&~ \partial_k\left( - W^k[J_k]+J_k\phivev\right) ~=~ -\left.(\partial_k W^k[J])\right|_{J=J_k}-\left.\left(\frac{\delta W^k[J]}{\delta J}\right)\right|_{J=J_k}\partial_kJ_k+\phivev \partial_kJ_k\,,\nonumber\\
~&=&~ - \left. (\partial_k W^k[J])\right|_{J=J_k} ~=~ -  \left. \frac{e^{-\intdx{  V_I\left(\frac{\delta}{\delta J}\right)}{} } \partial_k Z_0^k[J]}{ Z^k[J]}\right|_{J=J_k} \,,
\eea
where between the first and the second line we used the fact that $\delta W^k/\delta J[J^k]\equiv\phivev$, which is equivalent to Eqs.\eqref{eq:phik} and \eqref{eq:derivs}.
Let us clarify the physical meaning of this equation. On the left-hand side we see the $k$-derivative of the effective action $\Gamma^k$, written as a functional of an {\it arbitrary} classical field $\phivev$. This is the form that we want for our $k$ derivative because it encodes the scaling of the couplings in $\Gamma^k$ and removes any $k$-dependence in the field VEV which occurs when the current is kept fixed. On the right-hand side, the $k$-dependent current appears that would be required in the coarse-grained theory in order to produce this effective action with that classical field. }

To evaluate Eq.~\eqref{eq:flow} it is useful to define the $k$-derivative of the integral of the regulated current-current operator,
\beq
\label{eq:KK}
K^k[J]~=~\frac{1}{2}\int \frac{\dd^4 p}{(2\pi)^4} 
\int_0^\infty \dd t\ \partial_k \calG^k(t) \tilde J(-p)\tilde J(p) 
e^{-t(p^2+m^2)}~,
\eeq
so that by Eq.~\eqref{eq:JJ} we find that the scale-derivative in Eq.~\eqref{eq:flow} can be written \begin{eqnarray}
\partial_k Z_0^k[J]&~=~&
K^k[J]Z_0^k[J]\,.
\end{eqnarray}
This quantity is, of course, useful for deriving a flow equation because we are interested in the theory's response to changes in the cut-off $k$. Taking the $k$-derivative implies that $K^k$ can receive contributions only from the shell at $t\approx k^{-2}$ where the regulator is ``turning on''. Indeed, for the simplified example in the introduction where $\calG^k$ is a Heaviside function, the $k$-derivative is proportional to  $\delta(k^{-2}-t)$ and the $t$ integral becomes trivial to perform so that $K^k[J]$ is manifestly insensitive to any UV or IR divergences (as alluded to in the Introduction). 

{Inserting this expression into Eq.~\eqref{eq:flow} and rearranging then yields the following progenitor flow equation: 
\be
\label{eq:flow2}
\partial_k \Gamma^k[\phivev] ~=~ - \left.(\partial_k W^k[J])\right|_{J=J_k} ~=~ -  \left.\frac{e^{-
\intdx{V_I\left(\frac{\delta}{\delta J}\right)}{}} K^k[J] e^{\intdx{  V_I\left(\frac{\delta}{\delta J}\right)}{} } Z^k[J]}{ Z^k[J]}\right|_{J=J_k}\,.
\ee}
It is important to realize that this equation is exact and, moreover, it relates the effective action on the left to the {\it full} partition function (not just the quadratic part).  Our task is essentially to compute and understand the right-hand side of this equation. 
Thus, it is our ``master equation'' to determine the flow.  

To evaluate it entails a straightforward, although rather tedious, exercise in determining the non-trivial commutation relations between $\exp {[ -V_I (\delta/\delta J)]}$ and $K^k[J]$ factors.  
 We cover the steps in Appendix~\ref{app:result}, which finally yields 
{\begin{eqnarray}\label{eq:flow3}
\partial_k \Gamma^k[\phivev] 
&~=~&
\intdx{
\left(\frac{\delta}{\delta J(x)}\right)^2 K^k[J] \,V_I''\left(\phik(x)\right) 
}{x}
~+~ \intdx{ \frac{\delta}{\delta J(x)} K^k[J]\,V_I'\left(\phik(x)\right) }{x}\nonumber\\
&&~-~
\intdx{ 
\frac{1}{2}
\left(\frac{\delta}{\delta J(y)}\frac{\delta}{\delta J(x)} K^k[J]\right)V_I'\left(\phik(x)\right)V_I'\left(\phik(y)\right) 
}{x,y}
- K^k[J]~,
\label{eq:appresult}
\end{eqnarray}
}
where the primes indicate differentiation with respect to $\phi$, {and with the understanding that we must set 
$J\to J_k$ over the entire right-hand side of this expression. Once we do this, $\phik$ is replaced everywhere by $\phivev$, with $J$ taking the correct value of $J_k$. }

\subsection{A background field approach}
A way to simplify this flow equation is to define the effective action using the background field method~\cite{Abbott:1981ke}. Following Ref.~\cite{Scharnhorst:1987cx},one can first define the partition function of a quantum fluctuation $Q$ around the classical background $\bar \phi$ as
\be\label{eq:partitionback}
\tilde Z[J;\bar\phi]~=~N\int\mathcal DQ\exp\left\{-
\intdx{
S(\bar \phi+Q)}{}
+\intdx{ JQ}{}
\right\}~,
\ee
redefine the current as
\be
J'(x) = J(x) - \frac{\delta S}{\delta\bar\phi(x)}\,,
\ee
and show that the partition function of Eq.~\eqref{eq:partitionback} can be written as
\be\label{eq:fullpartition}
\tilde Z[J';\bar\phi]=e^{-S_\mathrm{I}\left[\frac{\delta}{\delta J'};\bar\phi\right]}\tilde Z_0[J';\bar\phi]\,,
\ee
where the interaction part of the classical action is implicitly defined through the expansion
\bea
S[\bar\phi+Q]=S[\bar\phi]+\int\dd^4x \ddelta{S[\bar\phi]}{\bar\phi(x)}Q(x)+\frac{1}{2}\int\dd^4x \int\dd^4y \frac{\delta^2S[\bar\phi]}{\delta\bar \phi(x)\delta\bar \phi(y)}Q(x)Q(y) + S_{\rm I}[Q;\bar\phi]\,,
\eea
and the partition function of the free theory is simply given by
\bea
\tilde Z_0[J';\bar\phi]&=& \tilde Z_0[0;\bar\phi]\exp\left\{-\frac{1}{2}\int\dd^4x J'(x)\left(\ddelta{{}^2S[\bar\phi]}{\bar \phi^2}\right)^{-1}J'(x)\right\}\,.
\eea
One can finally obtain the background effective action as
\be
\tilde \Gamma[\bar Q;\bar\phi]\equiv -\log \tilde Z[J';\bar\phi] + \intdx{J'\bar Q}{}\,,
\ee
where the classical value of the fluctuation $Q$ satisfies the equation of motion
\be
\bar Q = \frac{\delta \log Z[J';\bar\phi]}{\delta J'}\,.
\ee
By doing so, the effective action can be obtained by evaluating the background effective action for a vanishing fluctuation $\bar Q=0$ such that
\be
\Gamma[\bar\phi] = \tilde \Gamma[0,\bar \phi]\,,
\ee
which also corresponds to setting the source current $J'$ to zero, as
\be
J' = 0 \quad\Leftrightarrow\quad J = \frac{\delta S[\bar \phi]}{\delta \bar \phi}\,.
\ee
Following the same procedure as before using the background field method corresponds to performing the following substitution in the derivation above:
\bea
m^2\quad & \longrightarrow & \quad S^{(2)}\left[\bar\phi\right] \,,\nonumber\\
\intdx{V_I\left(\ddelta{}{J(x)}\right)}{x}\quad & \longrightarrow & \quad S_I\left[\ddelta{}{J};\bar\phi\right] = \intdx{\mathcal L_I\left[\ddelta{}{J(x)};\bar\phi\right]}{x}\,,
\eea
where in the last equality we assumed that we can consider the theory to be local, such that the interaction part of the background action can be written as the integral of a local Lagrangian interaction density $\mathcal L_I$, as
\be
S_I\left[Q;\bar\phi\right] = \intdx{\mathcal L_I\left[Q(x);\bar\phi(x)\right]}{x}\,.
\ee
Using the same procedure as described above to regularize the free partition function and obtain a regularized effective action yields the flow equation
\begin{eqnarray}\label{eq:flow4}
\left.\partial_k \tilde\Gamma^k[Q;\phivev]\right|_{Q=0} 
&~=~&
\intdx{\left.\frac{\delta^2K^k[J']}{\delta J'}\right|_{J'=0}  \,\left.\frac{\delta^2\mathcal L_I}{\delta Q(x)^2}\right|_{Q=0} }{x}
-\intdx{ 
\frac{1}{2}
\left.\left(\frac{\delta}{\delta J'(y)}\frac{\delta}{\delta J'(x)} K^k[J']\right)\right|_{J'=0} \left.\frac{\delta\mathcal L_I}{\delta Q(x)}\right|_{Q=0}\left.\frac{\delta\mathcal L_I}{\delta Q(y)}\right|_{Q=0}  
}{x,y}~.\nonumber\\
\end{eqnarray}
In this expression, we have dropped terms linear in $J'=0$ when enforcing that $Q=0$. Note, though, that the right hand side of this expression actually cancels exactly. Indeed, the interactions contained in the Lagrangian $\mathcal L_I$ are by definition at least trilinear, and evaluating first or second derivatives of this Lagrangian for vanishing fluctuations leads to a null result. This is an artefact of the background field method approach, which led us to absorb the $\exp(S[\bar\phi])$ contribution into the partition function in the normalisation factor, effectively considering the background as purely classical, and constituting the true vacuum of a quantum theory that lives on top of it. Extracting a meaningful flow equation thus requires taking higher derivatives of this equation, leading to 
\be
\partial_k \left(\frac{\delta\Gamma^k}{\delta \bar \phi(x)}\right) 
~=~\left.\partial_k \left(\frac{\delta\tilde\Gamma^k}{\delta \bar \phi(x)}[Q;\bar\phi]\right) \right|_{Q=0}
~=~\left.\partial_k \left(\frac{\delta\tilde\Gamma^k}{\delta Q(x)}[Q;\bar\phi]\right) \right|_{Q=0}
~=~
\left.\frac{\delta^2K^k[J']}{\delta J'}\right|_{J'=0}  \,\left.\frac{\delta^3\mathcal L_I}{\delta Q(x)^3}\right|_{Q=0}~.\label{eq:flow5}
\ee
Because the interacting part of the action is at least cubic in $Q$, we can as well use the fact that
\be
\left.\frac{\delta^3\mathcal L_I}{\delta Q(x)^3}\right|_{Q=0} = \left.\frac{\delta^3\mathcal S_I[Q;\bar\phi]}{\delta Q(x)^3}\right|_{Q=0} =  S^{(3)}[\bar \phi]\,,
\ee
such that the flow equation now becomes
\be
\partial_k \left(\frac{\delta\Gamma^k}{\delta \bar \phi(x)}\right) 
~=~\left.\frac{\delta^2K^k[J']}{\delta J'}\right|_{J'=0}  \,S^{(3)}[\bar \phi]~.\label{eq:appresultprelim}
\ee
Substituting the actual expression for $K^k[J']$, as derived in Appendix \ref{app:result} finally gives the flow equation
\be
\partial_k \left(\frac{\delta\Gamma^k}{\delta \bar \phi(x)}\right) 
~=~ S^{(3)}[\bar \phi]\int \frac{\dd^4p}{(2\pi)^4}\int_0^\infty {\dd t}\ \mathcal \partial_k \calG^k(t)\,  e^{-t\ (S^{(2)}[\bar \phi]+p^2)}\,.\label{eq:appresultprelim}
\ee

Until now, the analysis has been an application of field theory with the IR-regulated propagator in Eq.~\eqref{eq:Schwing_prop}. However, at this point, we reach the crux of our argument. In its present form, the result in Eq.~\eqref{eq:appresultprelim} is not useful for our purposes because the expression on the left is in terms of the effective theory, coarse-grained on scale $k^{-1}$, while the one on the right involves the bare theory, supposedly defined at $k\to \infty$ or some UV scale $k=\Lambda$ at which the effective field theory is known. The crucial last step in the procedure is therefore to take this exact equation for $\partial_k(\delta{\Gamma^k}/\delta\bar\phi)[\phivev]$, and deduce from it a corresponding flow equation in terms of the effective theory whose potential is $U^k(\phivev)$ rather than $V_I(\bar\phi_k)$. 


\subsection{Block Spin Coarse-Graining}

In order to obtain a flow equation that entirely refers to the theory averaged on scale $k^{-1}$, we adapt the central approach of Ref.~\cite{WETTERICH1991529}. The essential argument of 
Ref.~\cite{WETTERICH1991529}
as it applies in the current context is that, at a given $k$, we should be able to start from scratch with an entirely new field theory that has no regulator, but whose action is given by $\Gamma^k$, with
the effect of the regulator being absorbed into a set of new couplings defined at scale $k$. 
To get from the full path integral to such an expression that involves only the effective action, we first note that the effect of the regulator is to divide the modes into short- and long-wavelengths. Only the short-wavelength modes have been included in the regulated path integral, while the long-wavelength modes belong to our effective theory. Thus, the complete path integral can be recovered by a path integration of the $k$-regulated theory over the {\it remaining} long-wavelength fluctuations of the $k$-averaged fields $\bar \phi_k$ which have not yet been included. To incorporate these additional fluctuations we can promote $\bar \phi_k$ to a quantum field $\phi_k = \bar \phi_k+Q_k$, and write the full path integral as
\begin{align}
Z[J] &~=~ ~N\int\mathcal D  \phi \, \exp\left\{-S[\phi]+\intdx{ J \phi}{} \right\} 
 ~=~ N\int\mathcal D \phi_k \, Z^k [J]~,\end{align}
 where by Eq.~\eqref{eq:leg} we have 
 \beq 
Z^k[J] ~=~ \exp \left\{ -\Gamma^k[\phi_k]+[J\phi_k]~\right\}~,
 \eeq
and where ${\cal D}\phi_k$ implies 
that we should perform the path integral 
only over the modes with wavelengths longer than $k^{-1}$. 
Obviously, in the limit that $k\to 0$ there is nothing left to integrate over and $Z^k$ becomes simply $Z^k \to \exp(-\Gamma[\bar\phi])$. 
 
However we note also that we could instead have substituted $Z^k = e^{-W^k} $ from Eq.~\eqref{eq:GammaDef}, and traded $(J,\bar\phi_k+Q_k)$ for $(J^k,\phi=\bar\phi+Q)$ as in Eq.~\eqref{eq:derivs}. This gives  nothing other than the partition function of a new effective theory defined at scale $k$ where the cut-off is now associated with the current:
\begin{align}
Z ~&=~ ~N\int\mathcal D  \phi \,\exp\left\{-\Gamma^k[\phi]+\intdx{ J ^k\phi}{} \right\}
\,,
\label{eq:ident}
\end{align}

Using a block spin approach, we can now coarse-grain this effective theory averaged on scale $k^{-1}$ on a slightly larger scale, ${k'}^{-1}\equiv(k-dk)^{-1}$, and send $dk$ to zero to obtain a fully consistent flow equations with quantities exclusively defined at $k^{-1}$. Let us thus start with the partition function, as written in Eq.~\eqref{eq:ident}, and follow all the same steps of the original regularization procedure described above. In doing so, one obtains a flow equation by varying the regulated effective action with respect to the mode $k'$ as before, but now using $\Gamma_k$ as the action that encodes the nature of the interactions of the field theory at the initial scale $k$. What was the classical action $S$ now becomes $\Gamma^k$, and the flow equation trivially becomes

\begin{figure}
    \centering
    \includegraphics[width=0.6\linewidth]{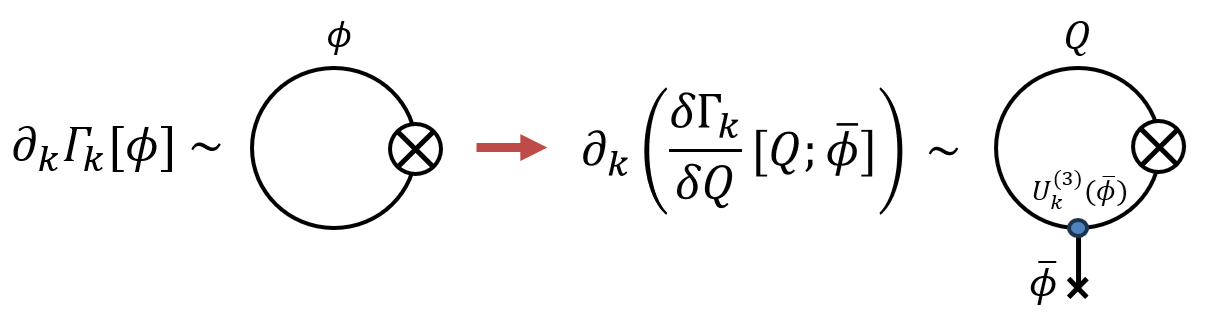}
    \caption{\label{fig:loop} Qualitative difference at the diagrammatic level between traditional {\em one-loop improved} ERG flow equations on the left and our background-field based flow equation on the right.}
\end{figure}

\begin{center}
\vspace{0.1cm}\fbox{\begin{minipage}
{0.9\linewidth}
\vspace{-0.1cm}
\be
\partial_k \left({\Gamma^k}^{(1)}[\bar\phi]\right) 
~=~ {\Gamma^k}^{(3)}[\bar \phi]\int \frac{\dd^4p}{(2\pi)^4}\int_0^\infty {\dd t}\ \mathcal \partial_k \calG^k(t)\,  e^{-t\ ({\Gamma^k}^{(2)}[\bar \phi]+p^2)}\,.\label{eq:RGeqGamma}
\ee
\vspace{-0.15cm}
\end{minipage}}~
\end{center}
{\vspace{0.2cm}}
This is our main result. Note that, as we have already discussed, it is only possible to derive flow equations for higher derivatives of the effective action in the background of the average field. The diagrammatic interpretation of the flow equations is therefore as illustrated in Fig.~\ref{fig:loop}: only the flow equation for couplings in the presence of a particular background field value can be deduced.

It will be illustrative to consider the flow equation for the corresponding effective potential in a concrete example. To do this we will focus on the ``local potential approximation'' (LPA). The LPA implies that  the effective potential at scale $k$, that is $U^k(\phi ) $, is given by projecting the functional $\Gamma^k[\phi]$ onto functions of homogeneous fields $\phi$ as 
\be
\Gamma^k[\phi ] ~=~ (\mathrm{Vol})~U^k(\phi) +\ldots \,,
\ee
where $U^k$ is the full potential (including the mass-squared term), while the ellipsis indicates kinetic terms and higher-order derivative terms. (The LPA is tantamount to the assumption that the kinetic terms remain canonically normalized.)
Using the LPA, this flow equation becomes
\begin{eqnarray}\label{eq:flow_potential}
\partial_k {U^k}^\prime\left(\bar \phi\right) &~=~& \frac{{U^k}^{\prime\prime\prime }(\bar \phi)}{32\pi^2}\int_0^\infty \frac{\dd t}{t^2}\ \mathcal \partial_k \calG^k(t)\,  e^{-t\ {U^k}''(\bar \phi)}\,.
\label{eq:RGeq}
\end{eqnarray}
Note that
although this expression is essentially the $\phi $ derivative of the one-loop inspired proposal in Ref.~\cite{Liao:1994fp} it was carefully derived from the properties of the all orders path integral, and at no point was it necessary to restrict our discussion to the one-loop effective potential.
This is indeed one of the main points of this paper. Note also that whereas Eq.~\eqref{eq:flow3} reproduces the flow in the LPA, it is completely general for scalar field theory.
For example, it would be possible to adopt more complete approximations than the LPA using this final result
as a starting point, for example, the so-called LPA$'$, which allows for adjustment due to field renormalization \cite{Dupuis:2020fhh}.

Finally, adopting for the moment the aforementioned Heaviside form for $\calG^k$, namely
$\calG^k = \theta (1- t k^{2})$, such that  $\partial_k\calG^k = -2 k t \delta(1- t k^{2})$, we find  
\begin{eqnarray}
\partial_k {U^k}^\prime\left(\bar \phi\right) &~=~& - k \frac{{U^k}^{\prime\prime\prime }(\bar \phi)}{16\pi^2} \,  e^{ - {U^k}''(\bar \phi)/k^2}\,~.
\label{eq:RGeq2}
\end{eqnarray}
This is the all-orders LPA  with a hard Schwinger-time cut-off function. {
It is worth noting before continuing that the effective action $\Gamma[\phi]$ must be convex because it is the Legendre transform of $\log Z[J]$. However,  $\Gamma^k[\phi]$  as defined in Eq.~\eqref{eq:ident} is {\it not} the Legendre transform of $\log Z^k[J^k]$, and thus the potential $U^k(\phi)$ need not be convex, which is of course in accord with our intuition about what should happen to the potential of the effective theory. }

\section{Comparison with standard momentum cut-off}

\label{sec:comparison}

\begin{figure}
    \includegraphics[keepaspectratio, width=0.49\textwidth]{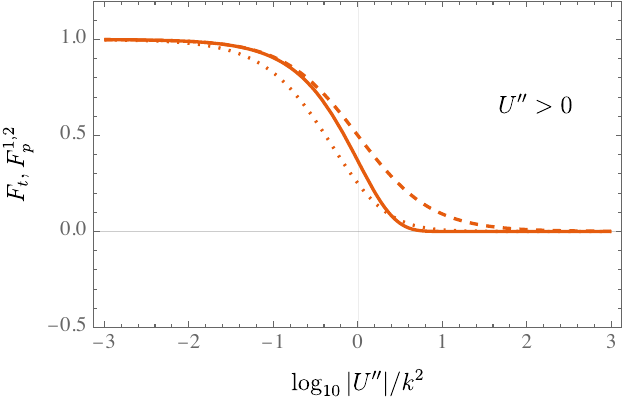}
   \includegraphics[keepaspectratio, width=0.49\textwidth]{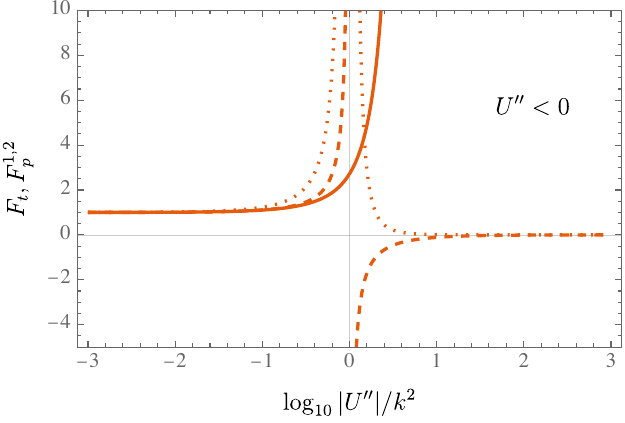}
    \caption{\footnotesize Schwinger regularized PTRG flow (solid line) versus canonical momentum cut-off flow (dashed line) and exact FRG (dotted line), where $\partial_k {U^k}' = - k \ {U^k}''' F({U^k}''/k^2)/16\pi^2 $, for the three possible flow functions $F_t,F_p^{1,2}$ discussed in the text. For a fixed value of ${U^k}''\equiv {U^k}''(\phi)$, when ${U^k}''(\phi)>0$ (left panel) the flows are similar. When ${U^k}''(\phi)<0$ (right panel) the Schwinger cut-off flow can be followed to $k^2<|{U^k}''|$, thus as $k\to 0$ freezing can be numerically established, while the momentum cut-off flow hits a singularity at $|{U^k}''|=k^2$. }
    \label{fig:comparison}
\end{figure}

Despite the superficial similarity to the standard ERG approach, there is a crucial difference, which will become important for our discussion of phase transitions. To elucidate it, let us now compare our flow equation to the canonical flow equation coming from a momentum regulator (see, e.g. Ref.~\cite{Berges:2000ew}). Under the same LPA approximation, that result is given by
\be
\label{eq:norm1}
\partial_k U_1^k ~=~ -\frac{k^3}{16\pi^2}\ln\left(1+\frac{{U^k}''}{k^2}\right)
\,.\ee
Alternatively, the exact RG flow equation derived in~Ref.~\cite{Wetterich:1992yh} with the optimized regulator $R_k = (k^2-q^2)\Theta(k^2-q^2)$ proposed in Ref.~\cite{Litim:2001up} leads to a flow equation of the form
\begin{equation}
\label{eq:norm2}
\partial_k U_2^k ~=~ \frac{k^3}{16\pi^2}\frac{1}{1+\frac{{U^k}''}{k^2}}\,.
\end{equation}
In our case, our results are nicely expressed as a flow equation for $U_\mu'$, so to make a direct comparison let us use the field derivative of Eq.~\eqref{eq:norm1} and Eq.~\eqref{eq:norm2}:
\begin{equation}
\partial_k {U_1^k}' ~=~ -\frac{k}{16\pi^2}{ {U^k}'''} ~F_{p}^1 \left( \frac{ {U^k}''}{k^2} \right) ~~;~~~~F_{p}^1(x) ~=~ \frac{1}{1+x}~.
\end{equation}
and
\begin{equation}
\partial_k {U_2^k}' ~=~ -\frac{k}{16\pi^2}{ {U^k}'''} ~F_{p}^2 \left( \frac{ {U^k}''}{k^2} \right) ~~;~~~~F_{p}^2(x) ~=~ \frac{1}{(1+x)^2}~.
\end{equation}
To compare with our Schwinger regulated flow equation  in Eq.~\eqref{eq:RGeq} let us continue with the Heaviside regulator such that  
\begin{equation}
\partial_k {U^k}' ~=~ -\frac{k}{16\pi^2}{ {U^k}'''} ~F_{t} \left( \frac{ {U^k}''}{k^2} \right) ~~;~~~~F_{t}(x) ~=~ e^{-x}~.
\end{equation}
To compare these equations, it is sufficient to compare the evolution of the three functions $F_t$ and $F_p^{1,2}$. 
These are shown in  Fig.~\ref{fig:comparison} for different signs of the mass term. 
The left panel shows the situation when ${U^k}''>0$. For a given positive ${U^k}''$, we see that decreasing $k$ causes a similar quenching for the Schwinger regulator (solid line), the more conventional momentum regulator (dashed line), and the ERG case with optimized regulator (dotted line). The crucial difference in behaviour arises when ${U^k}''<0$, however. Such regions clearly cannot be followed to low $k$ using the momentum regulator, because for negative curvature regions the functions $F_p^{1,2}$ diverge at $k^2=|{U^k}''|$. This does not necessarily indicate that there is a physical problem as recall we are plotting the coefficient of $\partial_k U'$. Hence, large positive values of $F_p$ could simply mean that the evolution completely freezes for these points with negative ${U^k}''$ once $k^2$ hits $|{U^k}''|$. However, this question is hard to resolve numerically with the conventional regulation procedure. Moreover, below this scale, the contributions are complex and difficult to interpret. Therefore, one usually simply declares the running frozen at the scale $k^2\approx |{U^k}''|$ in order to for example estimate the tunnelling rate. However,  
it is not clear whether the potential actually freezes at this scale or not. 

The Schwinger proper-time regulator, however, does not {\it necessarily} suffer from this problem as we shall see generically in the next section. Indeed in the above Heaviside  example the potential remains finite  at $k^2\approx |{U^k}''|$ and does not have a singularity, but continues to grow exponentially but smoothly as $k^2$ drops below $|{U^k}''|$, which allows one to follow the evolution into this region. This possibility is the most striking qualitative difference between the PTRG flow and that of the two momentum-based flows. Physically, the conclusion should be the same with all three flows we should add: the running freezes. However, the crucial point is that the running can freeze before or after the potential becomes convex. Indeed, this question is very subtle, and in order to determine the behaviour using the exact momentum cut-off result (instead of stopping the running by hand at $k^2\approx {U^k}''$) one has to adopt much more sophisticated numerical methods ~\cite{Alexandre:1998ts,Strumia:1998nf}.

\section{Dependence on the regulator profile} \label{sec:regulator}

Given the discussion above, it is worth determining which forms of proper time regulator allow us to pass through the point $U^{k\prime\prime} + k^2 =0$. The generic integral that we are required to perform 
is 
\begin{eqnarray}
\partial_k {U^k}^\prime\left(\bar \phi\right) &~=~& \frac{{U^k}^{\prime\prime\prime }(\bar \phi)}{32\pi^2}\int_0^\infty \frac{\dd t}{t^2}\ \mathcal \partial_k \calG^k(t)\,  e^{-t\ {U^k}''(\bar \phi)}\,.
\label{eq:RGeq22}
\end{eqnarray}
Most proper time flows that have been considered in the literature (see Ref.~\cite{Litim:2002xm}) have been motivated by the assumption that the IR behaviour of the cut-off should be controlled by $k$ acting as a mass, and have therefore taken $\partial _k \calG^k \propto e^{-k^2 t}$. 
Such flows can be represented as a linear expansion of the regulator in a basis of derivatives of the incomplete Gamma function: 
\begin{equation}
    \partial_k \calG_m^k(t) ~\propto~ k^{-1}(k^2 t)^m e^{-k^2 t}~. 
    \end{equation}
However these individual terms when inserted into the integral generate poles of the form
\begin{equation}
    \partial_k {U^k}^\prime\left(\bar \phi\right)~\propto \frac{1}{(k^2 + {U^k}'')^{m+1}}~,
\end{equation}
which obstruct the flow at $k^2 = -{U^k}''$.
It is easy to see what goes wrong: the Schwinger integral with this form of regulator is no longer uniformly convergent  once $k^2 + {U^k}''\leq 0$. However our procedure is simply based on a regulation of the entire partition function which suggests that these unphysical poles should be removed by using regulating functions that maintain uniform convergence for all $k$, which in turn requires  them to be of sufficiently rapid decay as $t\to\infty$.

For convenience let us define $\hat t = k^2 t $. Without loss of generality our regulator function can be written as a function of $k^2 t$, and furthermore its derivative can be written  \begin{equation}
    \partial_k \calG^k ~=~ -2 k t \, e^{-R(k^2 t)} ~=~ -2\frac{\hat t }{k} e^{-R(\hat t  )}~,
    \end{equation} 
and is peaked around $\hat t =1$. Our previous Heaviside function clearly corresponds to the highly peaked case, $e^{-R(\hat t )}\to \delta(\hat t -1 )$. More generally one might imagine a function like $R(\hat t) = \frac{(\hat t -1)^2 }{\sigma^2 } + \log \sigma\sqrt{\pi}  $ which would correspond (in the $\sigma\to 0$ limit) to the Gauss-Weierstrass form of the delta function. Note that since $\partial_k \calG^k (\hat t) = 2 k^{-1} \hat t \partial_{\hat t} \calG^k (\hat t)$ we must have 
\begin{equation}
    \int \frac{d\hat t}{\hat t} \partial _ k \calG^k (\hat t) ~=~ - 2 k^{-1}~.
\end{equation}
such that 
\begin{equation}
    \int_0^\infty {d\hat t} \, e^{-R(\hat t)} ~=~ - 1~.\label{eq:norm}
\end{equation}

We see the troublesome poles are avoided as long as the regulator functions displays exponentially damped behaviour at large $\hat t$,
\begin{equation}
    \lim _{\hat t\to\infty} \frac{\hat t}{R(\hat  t)} ~=~ 0~.
    \label{eq:scale}
\end{equation}
If this condition is satisfied then in the regime where ${U^k}''\ll 0$ the integral is dominated by a saddle  point at 
\begin{equation}
    \hat t_* ~\sim~ 1~.
\end{equation}
For example a generating function behaving asymptotically as $R(\hat t ) \sim ( \hat t )^m $ yields $\hat t_* =  \left(\frac{|{U^k}''|}{mk^2}\right)^{\frac{1}{m-1}}$ and hence an answer which is finite for negative $U^{k\prime\prime}$:
\begin{equation}
\partial_k U ^{k\prime} ~\propto ~\frac{{U^k}^{\prime\prime\prime }(\bar \phi)}{32\pi^2} \exp \left[ (m-1) \left( \frac{|{U^k}''|}{m k^2} \right)^{\frac{m}{m-1}}\right]~.
\end{equation}
More generally we find a saddle point at solutions to  $ R'(\hat t_*)= |{U^k}''|/k^2$. One might be concerned about the possibility of divergences at $k\to 0$. However as we shall see in the next section, the potential either freezes at finite $k$ in the LPA approximation, or eventually becomes  convex in the $k\to 0$ limit, removing this potential problem.

Within this class of regulators it is natural to ask if the precise form of the regulator profile is important. Indeed, the Heaviside function is somewhat special because it renders the integrals trivial. To consider this question, we note that, making the change of variables $\hat t = s+1$, 
 our integral in Eq.~\eqref{eq:RGeq22} can be written  
\begin{eqnarray}
\partial_k {U^k}^\prime\left(\bar \phi\right) &~=~& -\frac{{U^k}^{\prime\prime\prime }(\bar \phi)}{16\pi^2}k e^{- {U^k}''(\bar \phi)/k^2 }\int_{-1} ^\infty \frac{\dd s}{(s+1)}\ e^{-R(1+s) } \,  e^{-s\ {U^k}''(\bar \phi)/k^2 }
\label{eq:RGeq23}
\end{eqnarray}
The function $R(1+s)$ can be expanded around its peak at $s=1$. Expanding to second order in $s$ and imposing the normalisation condition in Eq.~\eqref{eq:norm} of course recovers precisely the Gauss-Weierstrass function 
\begin{equation}
   e^{-R}~\approx ~ \frac{1}{\sigma \sqrt{\pi}} e^{-  s^2/\sigma^2 } ~,
\end{equation}
where $\sigma \approx \sqrt{2/R''(1)}$. This gives 
\begin{eqnarray}
\partial_k {U^k}^\prime\left(\bar \phi\right) &~=~& -\frac{{U^k}^{\prime\prime\prime }(\bar \phi)}{16\pi^2}k e^{- {U^k}''(\bar \phi)/k^2 }
e^{ {\sigma^2}\,\frac{{U^k}''(\bar \phi)}{4 k^2}  }~,
\label{eq:RGeq24}
\end{eqnarray}
which becomes universal for ${\sigma^2{U^k}''(\bar \phi)}\ll { k^2}$. This 
tendency to universality is due to the normalisation and asymptotic conditions in Eq.~\eqref{eq:norm} and \eqref{eq:scale} respectively.  

Clearly the scale $\delta t \equiv \frac{1}{\Lambda^2 } =  \sigma^2/ k^2   $ is acting as the thickness of the shell that is being integrated out. Note that the ratio $k/\Lambda$ remains constant in this scheme for fixed $\sigma$, implying that the width of the regulator profile keeps a constant proportion with respect to the regulator function as a whole as $k$ varies. In such a scheme there is no limit on the value of $k$ in the UV, although a UV cut-off would still of course be required to regulate UV divergences.  In this context it is worth remarking that in UV complete settings, in order to maintain UV-completeness $\calG$ must be a function that is itself built out of fundamental degrees of freedom~\cite{Abel:2021tyt,Abel:2023hkk}. Thus the renormalisation scale $k^2$ and the width of the cut-off $\sigma^2 /k^2$ are built from a single dimensionful parameter (for example a compactification radius).  Thus fixed $\sigma$  has a well defined operational meaning, while the scale $\Lambda$ is naturally related to the fundamental scale.

\section{Towards Convexity}
\label{sec:towards}

When going from the classical limit at $k\to \infty$ to the fully renormalised effective potential at $k\to 0$, one can study the analytical form of our flow equation to see whether the potential evolves towards convexity. A convex function features a second derivative that is positive across its domain of definition. When considering a locally non-convex potential, one can thus define a finite domain \be
\mathcal D^k_-~\equiv~\left\{\phi \ |\ {U^k}''(\phi)<0\right\}~.
\ee
A function that evolves towards convexity would typically see this domain shrink for decreasing $k$, and reduce to the empty set once the function becomes completely convex. 

To see whether our flow equation acts on the effective potential in this manner, we define the field value that minimizes the potential's {\it second derivative} as
\be
\phi_{\rm min}^k ~\equiv ~\min_{\phi\in \mathcal D_-^k} \left\{{U^k}''(\phi)\right\}~.
\ee
This is the point where the potential is locally  ``maximally concave''.
By definition, if it is concave at this point, then the potential has the following properties:
\be
\label{eq:properties}
{U^k}''(\phi_{\rm min}^k)~<~0\,,\quad {U^{k}}'''(\phi_{\rm min}^k)~=~0\,,\quad\text{and}\quad {U^{k}}''''(\phi_{\rm min}^k)~>~0\,.
\ee
Let us now determine how the potential's second derivative evaluated at this minimum evolves. Using the properties in Eq.~\eqref{eq:properties} together with the flow equation derived in Eq.~\eqref{eq:flow_potential} we find the following behaviour for the scale dependence of the curvature at the maximally concave point:
\begin{eqnarray}
\frac{\dd}{\dd k}\left[{U^k}''(\phi_{\rm min}^k)\right] &~=~& \frac{\dd\phi_{\rm min}^k}{\dd k}\underbrace{{U^{k}}'''(\phi_{\rm min}^k)}_{=0}\,+~\partial_k {U^k}''(\phi_{\rm min}^k) ~, \nonumber\\
&~=~& \frac{1}{32\pi^2}\left[\underbrace{{U^{k}}''''(\phi_{\rm min}^k)}_{>0}\int\frac{\dd t}{t^2}\partial_k \mathcal G^k(t) e^{-t {U^k}''(\phi_{\rm min}^k)} ~-~\left(\underbrace{{U^{k}}'''(\phi_{\rm min}^k)}_{=0}\right)^2\int\frac{\dd t}{t}\partial_k \mathcal G^k(t) e^{-t {U^k}''(\phi_{\rm min}^k)}\right]~,\nonumber\\
&~=~ & 
- \frac{k}{16\pi^2}
\underbrace{{U^{k}}''''(\phi_{\rm min}^k)}_{>0} e^{- {U^k}''(\phi_{\rm min}^k)/k^2} \left( 1~+~\calO \left(\frac{k^2}{\Lambda^2}\right)\right)~,
\nn \\
&~<~&0~,
\label{eq:convex_deriv}
\end{eqnarray}
where in the last step we perform the same procedure that we used in Sec.~\ref{sec:regulator}, with $1/\Lambda^2$ being the width of the regulator profile.

This proves that if the effective potential is locally non-convex at the scale $k$, then the flow equation pushes its second derivative towards positive values as $k$ drops. As we anticipated in previous sections, this does not mean that the potential will necessarily reach complete convexity, because from Eq.~\eqref{eq:convex_deriv} it is clear that the flow equation is efficient only when $k^2\gg U''_k$. But nevertheless we see that the overall action of the flow is always in favour of convexity as $k$ decreases. 

Whether a scalar potential reaches convexity or not is a crucial question because non-convex potentials that feature a metastable minimum could lead to a first-order phase transition involving quantum tunnelling out of the false vacuum, whereas a 
convex potential could only ever describe second-order phase transitions. Thus, we would like to use the PTRG flow equations to determine whether the traditional tunnelling computation can even be valid and, if so, for what effective parameters. We now turn to this question.

\section{Quantum tunnelling}
\label{sec:tunnelling}

{
The study of phase transitions using functional renormalisation has emerged in a variety of scalar models, providing a reliable way to determine non-universal quantities (e.g. critical temperatures) and universal ones (e.g. critical exponents)~\cite{Tetradis:1992xd, Tetradis:1993bx, Tetradis:1993ts, Reuter:1993rm, Tetradis:1995br, Berges:1995mw, Bornholdt:1994rf, Bornholdt:1995rn, Bornholdt:1996ir}. In the context of gauge
theories~\cite{Reuter:1992uk, Reuter:1993nn, Reuter:1993kw, Reuter:1994sg}, ERG techniques have led to the study of second-order and first-order phase transitions for the Abelian and non-Abelian Higgs models~\cite{Litim:1994jd, Bergerhoff:1995zm, Bergerhoff:1995zq, Bergerhoff:1994sj, Bergerhoff:1995aa, Bergerhoff:1996jw, Tetradis:1996fw}, with implications for the electroweak phase transition~\cite{Tetradis:1996fw} and the chiral phase transition~\cite{Berges:1997eu, Berges:1996ib}. Applying ERG flows to the study of nucleation rates in first-order phase transitions was then studied in Refs~\cite{Berges:1996ib, Berges:1996ja, Strumia:1998nf, croon2021non}. In this section, we aim to illustrate how the use of a proper-time ERG flow can help alleviate technical obstacles that may arise in this context.}

\begin{figure}
    \centering \includegraphics[width=0.499\linewidth]{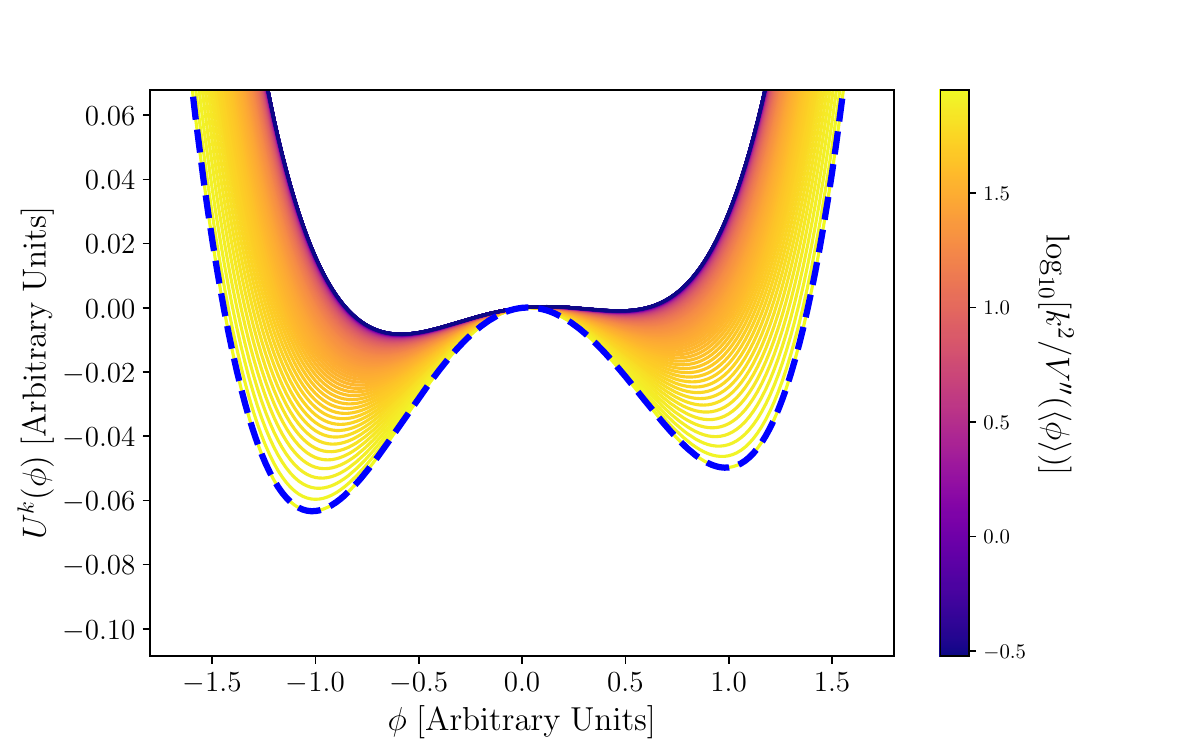}\includegraphics[width=0.499\linewidth]{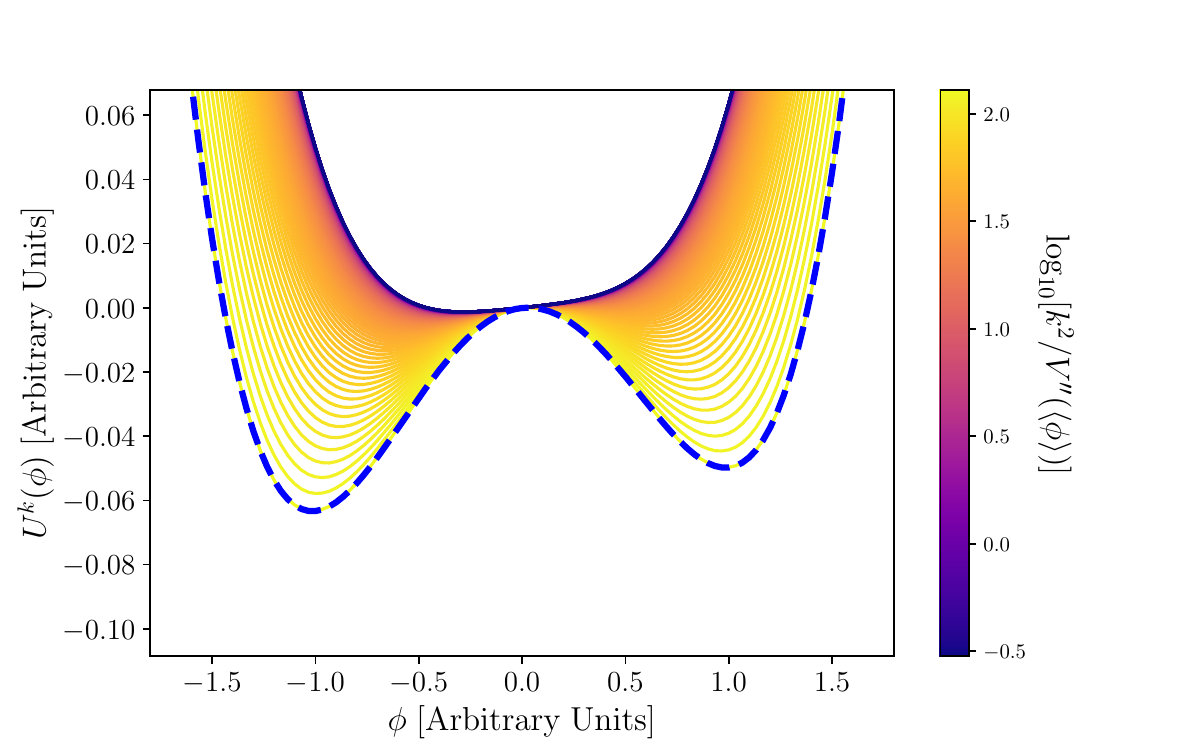}
    \caption{\label{fig:flow_num} \footnotesize Evolution of the effective potential $U^k(\phi)$ with the scale $k$, starting with an arbitrary potential $V(\phi)$ at scale $k_{\rm UV}$ (dashed blue line). In the left panel (right panel), $k_{\rm UV}^2 = 80\times m^2$ ($k_{\rm UV}^2 = 130\times  m^2$).}
    
\end{figure}
 
To show how one can apply PTRG flow equation to the problem of false-vacuum decay we will look at a particular example. Starting at an arbitrary UV scale $k=k_{\rm UV}$, we consider the following potential:
\be
V (\phi)\equiv \frac{\lambda}{4}\left( \phivev^2 - \phi^2\right)^2  + \frac{\epsilon }{2 \phivev}  \left(\phi-\phivev\right)\,. 
\ee
In this expression, $\phivev$ is the vacuum expectation value (vev) of the field in the true minimum, while $\epsilon $  is the potential difference between the metastable minimum and the true minimum. 
That is, the parameter $\epsilon$ controls the asymmetry of the potential with respect to the origin, and the combination $m\equiv \sqrt{\lambda}\phivev$ controls the mass of the field in the true vacuum. Thus, parameters can be chosen so as to make the potential more or less convex, with the false vacuum having accordingly varying tunnelling solutions at the starting scale $k_{\rm UV}$.

We track the evolution of this potential with the energy scale $k$ by using the flow equation~ \eqref{eq:flow_potential} with the regulator introduced in Sec.~\ref{sec:regulator} whose internal cut-off scale we choose to be $\Lambda\gg k_{\rm UV}$ such that our results are insensitive to that choice. We then propagate our flow equation starting from $U^k = V$ at $k=k_{\rm UV}$, obtaining the new shape of the potential at smaller values of $k$ step by step, by solving numerically the flow equation for ${U^k}'(\phi)$ and integrating numerically over $\phi$ to recover $U^k(\phi)$. In Fig.~\ref{fig:flow_num}, we exhibit two situations where we choose the same initial potential $V(\phi)$ but vary the initial scale $k_{\rm UV}$ such that the potential has more or less room to evolve before it freezes out. In the right panel of Fig.~\ref{fig:flow_num}, this leads to a potential that becomes completely convex in the IR, whereas in the left panel, the potential freezes while still not fully convex. 

It is only in the latter case that the traditional decay rate computation can be employed. Therefore, let us now turn to this particular aspect and discuss in more detail how we should identify and treat these two situations. To calculate the decay rate of a metastable (false) vacuum of a potential $U(\phi)$, one needs to determine the aforementioned tunnelling solution, which is an $O(4)$-symmetric `bounce' instantonic solution interpolating between the true vacuum at $\rho=0$ ($\rho$ the radial coordinate in euclidean space) and the false vacuum at infinity, while minimizing the Euclidean action~\cite{Coleman:1977py},
\begin{align}
\label{eq:S4}
S_4 ~&=~ \int d^4x ~\frac{1}{2} \dot\phi^2
+\frac{1}{2} (\nabla\phi)^2 + U(\phi)~,\nonumber \\
~&\equiv ~~ 2\pi^2 \int \rho^3 d\rho~~\frac{1}{2} (\partial_\rho\phi)^2 + U(\phi)~.
\end{align}
The decay rate per unit volume is then given by 
\beq
\Gamma ~=~ Ae^{-S_4}~.
\eeq
In this expression, the pre-exponential factor is given by 
\begin{equation}
A~=~
\left( \frac{S_4}{2\pi} \right)^2
\left(  
\frac
{\det' (-\square +U'' (\phi) ) } 
{\det (-\square +U'' (0) ) } 
\right) ^{-1/2}~,
\end{equation}
where $U(\phi(\rho))$ is the potential as a function of the instanton solution $\phi(\rho)$ and the prime indicates that the zero eigenvalues of $\square -U''(\phi)$ are to be omitted. 

Note that this overall structure comes from integrating out the small wavelength modes around the instanton solution corresponding to the theory with the $k$-dependent effective potential, and thus, such an integration should terminate at the scale $k$ above the effective field theory.
Consequently, the prefactor $A$ also has $k$ dependence in both $S_4$ and $U$.  However, we know that, if for example the length scale $1/k$ is much smaller than a bubble wall, then changing it should not affect the tunnelling rate, which is dominated by physical length scales larger than the width of the bubble walls. In other words, the separation into an instanton contribution and a fluctuation contribution has a spurious $k$ dependence that is not physical.  Thus, the important aspect of such computations for our purposes is that for them to make sense the $k$-dependence in the $A$ prefactor should in principle cancel the $k$-dependence in the exponential factor \cite{Berges:2000ew}.

Establishing this overall $k$-independence can be achieved with the kind of numerical analysis carried out in Ref.\cite{Berges:2000ew}. However, the present approach allows us to sidestep these difficulties somewhat because we are able to follow just the instanton solution to low $k$. If the solution freezes and becomes $k$-independent in the IR then we can be sure that the prefactor $A$ must be constant there as well. The system in this region represents a scale-independent potential that can be used to perform a canonical tunnelling calculation with an assumed constant prefactor. 

Thus rather than evaluating the full decay rate, we focus here on the evolution of the instanton solution with energy scale, as the effective potential heads towards convexity. 
Thus, the panels shown in Fig.~\ref{fig:flow_num} demonstrate the two possible situations that one has to deal with. The left panel where the potential freezes as a function of $k$ is the aforementioned situation, which would lead to an effective action in the IR that does feature a metastable minimum. In this case, the traditional computation is valid. Meanwhile, in the right panel the potential becomes convex, and a phase transition can only be second-order. Here, any conclusions based on a traditional tunnelling calculation would be spurious. 

\begin{figure}
    \centering
\includegraphics[width=0.75\linewidth]{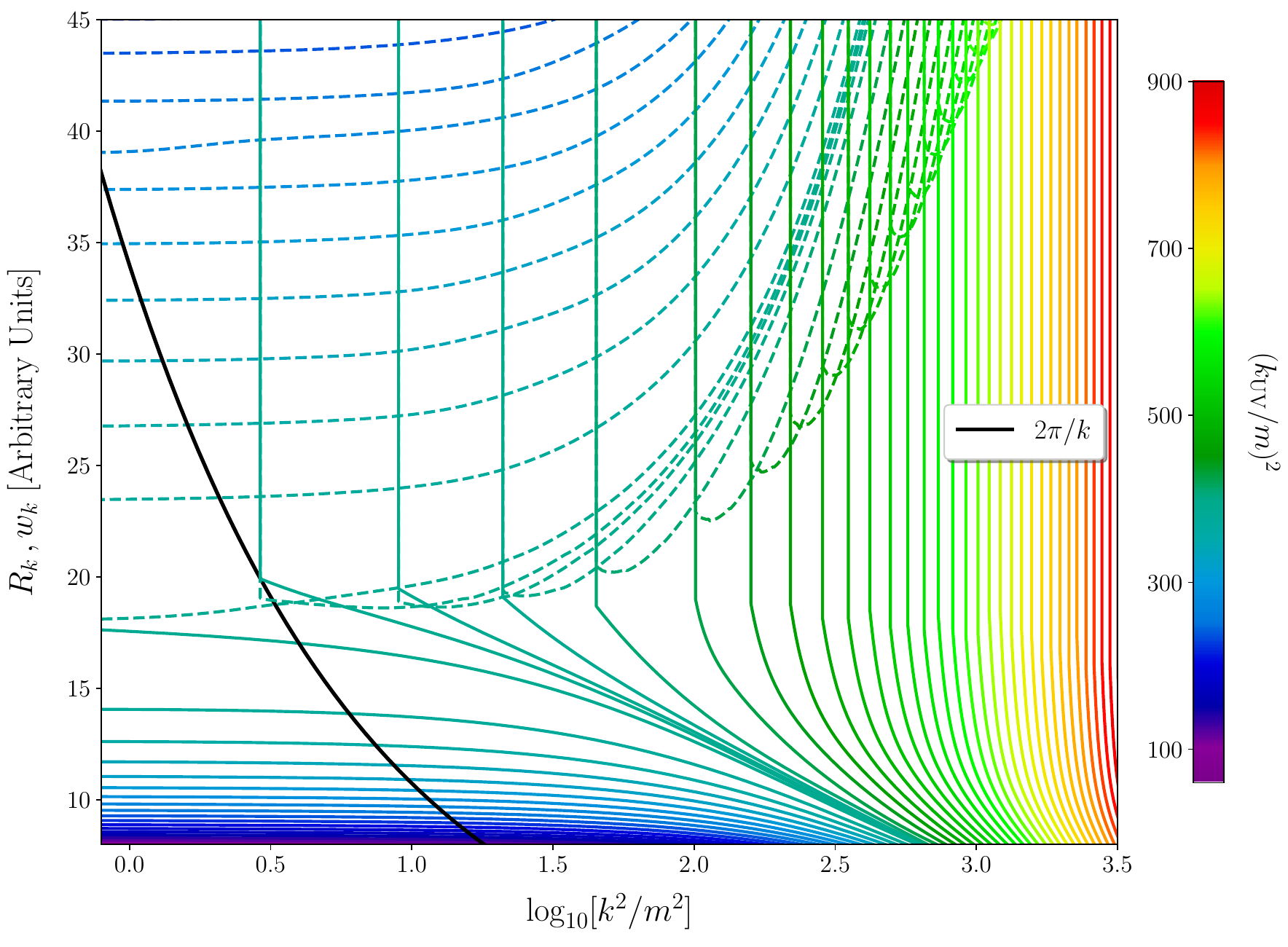}
\caption{\label{fig:bubble_final} \footnotesize Bubble configurations calculated for a potential $V(\phi)$ featuring $\phivev=1$, $\lambda\approx 0.6147$ and $\Delta V\approx 0.01817$, defined at the initial UV scale $k_{\rm UV}^2 \in [10^2\times m^2,10^3\times m^2]$. Dashed lines represent the bubble radius $R_k$, plain lines the bubble wall's width $w_k$, and the black line is the scale $2\pi/k$. We  can observe that this line separates the regimes where the potential becomes convex (upper region) and where the potential freezes while non-convex (lower region). When the potential becomes convex, the wall thickness and bubble radius merge in a thick-wall configuration, which rapidly becomes infinitely large as the potential becomes convex. Reaching a frozen thick-wall regime where $R_k\sim \omega_k$ requires a certain amount of tuning.  }
    
\end{figure}
To investigate the behaviour in further detail, let us now consider a potential $V(\phi)$ with a very small initial $\epsilon$ parameter, for which the bounce solution calculation interpolating between the false and the true vacuum describes a thin-wall bubble profile~\cite{Coleman:1977py}. At $k=k_{\rm UV}$, the width of the bubble wall, $w_k$, is thus much smaller than the overall radius of the bubble, $R_k$. Then, we track the evolution of the effective potential $U^k$ with the scale $k$ and calculate numerically the bounce solution to obtain the corresponding bubble profile. In Fig.~\ref{fig:bubble_final}, we plot this evolution (from right to left) for the bubble radius (dashed lines) and the bubble wall's width (plain lines) for a set of initial UV scales $k_{\rm UV}$ ranging from $10 \times m$ (purple lines) to $\sim 33 \times m$ (red lines). As one can see from the figure, all bubbles start at $k=k_{\rm UV}$ from a thin-wall configuration where $R_k\gg w_k$. When $k$ decreases, the bubble radius decreases, and its width increases as the flow equation slowly pushes away the metastable minimum. Eventually, after the potential freezes, the metastable minimum may either disappear (turning into an inflexion point first before the potential becomes convex) or survive the RG flow evolution. In the former case, the mass of the field in the false vacuum goes through zero and the bubble becomes thick-wall, which explains why the cyan-to-red coloured curves all grow for decreasing $k$ before the potential freezes\footnote{In practice, we stopped simulations when the local minimum disappears.}. In the latter case, the bubble profile freezes before the bubble configuration is thick-wall ($R_k\gtrsim w_k$). 

In Fig.~\ref{fig:bubble_final}, we also indicate the evolution of the coarse-graining length scale $2\pi/k$ for comparison. Remarkably, the potentials --- and the corresponding bubble configurations --- freeze when $w_k \sim 2\pi/k$. This is understandable, as the only modes contributing to the instanton solution are necessarily on scales smaller than the bubble width, and larger wavelengths do not affect physical processes on small scales. This highlights one of the key features of this approach for the calculation of a false-vacuum decay: unlike the case of a momentum cut-off that requires an additional cut-off on the field value to account for relevant modes when calculating a tunnelling rate~\cite{croon2021non}, one does not need to know the scale at which the scale $k^{-1}$ becomes larger than the bubble, as the potential naturally freezes there. Instead, to determine whether a phase transition is first-order or second-order, it is sufficient to run the PTRG equation down to $k=0$ and consider the obtained frozen potential.  If the potential is convex at this point, then the transition has to be second order. If it still possesses a metastable minimum, one can use the full effective potential to calculate the physical rate of vacuum decay via the usual bounce solution. Nevertheless, we can observe from Fig.~\ref{fig:bubble_final} that stopping the running when $\omega_k = 1/k$ gives an accurate indication of the true tunnelling potential as long as the one avoids the thick-wall regime. 

 As a final clarifying remark, we wish to emphasize the following point\footnote{We thank Djuna Croon for discussions on this point.}. We know that the full effective potential should be convex: what then is the {\it physical} ingredient in our analysis which allows the potential to freeze in a non-convex state as we send $k\to 0$? The answer lies in the LPA  that we used in order to perform the flow throughout the tunnelling study. The LPA assumes homogeneous fields. It is thus only good for describing a local EFT potential, and it excludes precisely those inhomogenous instantonic modes that are required to restore convexity in the event that the length scale $1/k$ becomes larger than the typical bubble wall size.

This completes our all-orders path integral derivation of the proper time renormalisation group (PTRG) equation and the discussion of its application to tunnelling using the LPA. The derivation is exact in the sense of Ref.~\cite{Litim:2001hk}, namely that it was derived from an all orders treatment of the deformed path integral. However the exact PTRG result differs from the more conventional ERG treatment of  Refs.~\cite{WETTERICH1991529,WETTERICH199390,Dupuis:2020fhh,Berges:2000ew},
in that it allows one to derive the flow of couplings in the background field formalism, but not of the entire effective action. This qualitative distinction is as shown in Fig.~\ref{fig:loop}.

We would like to emphasise that the techniques we have developed here can, in principle, be extended to cover effective actions that include more than one kind of particle. As a proof of principle, we derive in Appendix~\ref{app:fermions} the flow equation for a Lagrangian involving one scalar field and one Majorana fermion. 
As can be seen there, such cases can contain other terms in the flow equation with other mass scales in the exponential. 
If one performs a tunnelling study similar to the above for these more general cases, one might expect situations where the running does not quite freeze due to light states to which the scalar continues to couple. However, one can anticipate that there is still a transition from a rapidly running potential to a quasi-frozen one. We leave the study of generalised exact PTRG flow equations for future work.\\

\noindent{\bf Acknowledgements:} We are very grateful to Djuna Croon and Nick Tetradis for helpful discussions and for pointing out to us key references on the topic.  This work was supported by the STFC under grant ST/P001246/1. LH would like to acknowledge support by Institut Pascal at Université Paris-Saclay during the Paris-Saclay Astroparticle Symposium 2023, with the support of the P2IO Laboratory of Excellence (program “Investissements d’avenir” ANR-11-IDEX-0003-01 Paris-Saclay and ANR-10-LABX-0038), the P2I axis of the Graduate School of Physics of Université Paris-Saclay, as well as IJCLab, CEA, IAS, OSUPS, and the IN2P3 master projet UCMN.

\appendix

\section{Deriving the flow equation}

In this appendix we show how to 
evaluate the right hand side of the  flow equation in Eq.~\eqref{eq:flow2}, 
\be
\partial_k \Gamma^k[\phi] ~=~ - \partial_k \ln Z^k[J] = -  \frac{e^{-\intdx{  V\left(\frac{\delta}{\delta J}\right)}{}} K^k[J] e^{\intdx{  V\left(\frac{\delta}{\delta J}\right)}{}} Z^k[J]}{ Z^k[J]}\,.
\ee
Writing $K_k[J]$ in euclidean space coordinate, we obtain
\be
K_k[J] = \frac{1}{2}\left(\int \frac{\dd^4 p}{(2\pi)^4} \int_0^\infty \dd t\ \mathcal \partial_k G_k(t) \int \dd^4x\dd^4y J(x)J(y)e^{i(x-y)p} e^{-t(p^2+m^2)}\right)\,.
\ee
Because we are going to act on $K_k[J]$ with the operator $V\left(\frac{\delta}{\delta J}\right)$ , let us apply a few successive functional derivatives first:
\begin{eqnarray}
\frac{\delta}{\delta J(x)}K_k[J]&=&\int \frac{\dd^4 p}{(2\pi)^4} \int_0^\infty \dd t\ \mathcal \partial_k G_k(t) \int \dd^4y J(y)\cos \left[(x-y)p\right] e^{-t(p^2+m^2)}\,,
\end{eqnarray}
and
\begin{eqnarray}\label{eq:xy2nd}
\frac{\delta}{\delta J(x)}\frac{\delta}{\delta J(y)}K_{k}[J]&=&\int \frac{\dd^4 p}{(2\pi)^4} \int_0^\infty \dd t\ \mathcal \partial_k G_k(t) \cos \left[(x-y)p\right] e^{-t(p^2+m^2)}\,,
\end{eqnarray}
and by setting $x=y$ in the previous equation we obtain
\begin{eqnarray}
\left(\frac{\delta}{\delta J(x)}\right)^2K_k[J]&=&\int \frac{\dd^4 p}{(2\pi)^4} \int_0^\infty \dd t\ \mathcal \partial_k G_k(t) e^{-t(p^2+m^2)}\,.
\end{eqnarray}
Before we continue, let us also  integrate these quantities over $x$:
\begin{eqnarray}
\intdx{\frac{\delta}{\delta J(x)}K_k[J]}{x} &=&  \int_0^\infty \dd t\ \mathcal \partial_k G_k(t) \int \dd^4y J(y)\cos (yp) e^{-t m^2}\,,\nonumber\\
\intdx{\left(\frac{\delta}{\delta J(x)}\right)^2K_k[J]}{x} &=& (\mathrm{Vol})\int \frac{\dd^4 p}{(2\pi)^4} \int_0^\infty \dd t\ \mathcal \partial_k G_k(t) e^{-t(p^2+m^2)}\,.
\end{eqnarray}
Let us now compute the numerator of Eq.~\eqref{eq:flow2}. Let us denote by $\mathcal O[1]$ the operator
\be
\mathcal O[1] \equiv e^{-\intdx{  V\left(\frac{\delta}{\delta J}\right)}{}} K_k[J] e^{\intdx{  V\left(\frac{\delta}{\delta J}\right)}{}}\,, 
\ee
and generalize it to the class of operators
\be
\mathcal O[\lambda] \equiv e^{-\lambda\intdx{  V\left(\frac{\delta}{\delta J}\right)}{}} K_k[J] e^{\lambda\intdx{  V\left(\frac{\delta}{\delta J}\right)}{}}\,. 
\ee
We have
\be\label{eq:opevo}
\frac{\dd\mathcal O}{\dd\lambda} = \intdx{  e^{-\lambda\intdx{  V\left(\frac{\delta}{\delta J(y)}\right)}{y}} \left[-V\left(\frac{\delta}{\delta J(x)}\right),K_k[J]\right] e^{\lambda\intdx{  V\left(\frac{\delta}{\delta J(y)}\right)}{y}}}{x}\,.
\ee
Note that in this expression, the brackets without subscripts $[\cdot,\cdot]$ denote the usual commutator rather than a space integral. In order to compute this commutator, let us write down the potential as an infinite series
\be
V(\phi)\equiv \sum_{k=3}^\infty a_k \phi^{k}\,.
\ee
Applying the commutator to a test function $\varphi[J]$, we then can write
\begin{eqnarray}
&&\left[-\sum_{n=3}^\infty a_n \left(\frac{\delta}{\delta J(x)}\right)^{n}, K_k[J]\right]\varphi[J] \nonumber\\
&=& -\sum_{n=3}^\infty a_n \left[\left(\frac{\delta}{\delta J(x)}\right)^{n} K_k[J]\varphi[J]- K_k[J]\left(\frac{\delta}{\delta J(x)}\right)^{n}\varphi[J]\right]\,,\nonumber\\ 
&=&-\sum_{n=3}^\infty a_n \left[\left(\sum_{j=0}^n\begin{pmatrix}n\\j\end{pmatrix}\left(\frac{\delta}{\delta J(x)}\right)^{j} K_k[J]\left(\frac{\delta}{\delta J(x)}\right)^{n-j}\varphi[J]\right)- K_k[J]\left(\frac{\delta}{\delta J(x)}\right)^{n}\varphi[J]\right]\,,\nonumber\\
&=&-\sum_{n=3}^\infty a_n \left[\sum_{j=1}^k\begin{pmatrix}n\\j\end{pmatrix}\left(\frac{\delta}{\delta J(x)}\right)^{j} K_k[J]\left(\frac{\delta}{\delta J(x)}\right)^{n-j}\varphi[J]\right]\,,\nonumber\\
&=&-\sum_{n=3}^\infty a_n \left[\frac{n(n-1)}{2}\left(\frac{\delta}{\delta J(x)}\right)^2 K_k[J]\left(\frac{\delta}{\delta J(x)}\right)^{n-2}\varphi[J]+n\frac{\delta}{\delta J(x)} K_k[J]\left(\frac{\delta}{\delta J(x)}\right)^{n-1}\varphi[J]\right]\,,\nonumber\\
&=& -\left[\left(\frac{\delta}{\delta J(x)}\right)^2 \frac{K_k[J]}{2} V''\left(\frac{\delta}{\delta J(x)}\right)+\frac{\delta}{\delta J(x)} K_k[J]V'\left(\frac{\delta}{\delta J(x)}\right)\right]\varphi[J]\,.
\end{eqnarray}
This leads to
\begin{eqnarray}
\frac{\dd\mathcal O}{\dd\lambda} &=& \intdx{  e^{-\lambda\intdx{  V\left(\frac{\delta}{\delta J(y)}\right)}{y}} \left[-V\left(\frac{\delta}{\delta J(x)}\right),K_k[J]\right] e^{\lambda\intdx{  V\left(\frac{\delta}{\delta J(y)}\right)}{y}} }{x}\,,\nonumber\\
&=&-\intdx{  e^{-\lambda\intdx{  V\left(\frac{\delta}{\delta J(y)}\right)}{y}} \left[\left(\frac{\delta}{\delta J(x)}\right)^2 \frac{K_k[J]}{2} V''\left(\frac{\delta}{\delta J(x)}\right)+\frac{\delta}{\delta J(x)} K_k[J]V'\left(\frac{\delta}{\delta J(x)}\right)\right] e^{\lambda\intdx{  V\left(\frac{\delta}{\delta J(y)}\right)}{y}} }{x}\,.\nonumber\\
\end{eqnarray}
As we have seen the second derivative of $K_k[J]$ is $J$-independent, however, the first derivative is linear in $J$, which leads to the simplification
\begin{eqnarray}\label{eq:dOdlambda}
\frac{\dd\mathcal O}{\dd\lambda} &=& -\intdx{ \left(\frac{\delta}{\delta J(x)}\right)^2 \frac{K_k[J]}{2} V''\left(\frac{\delta}{\delta J(x)}\right)}{x}-\intdx{ e^{-\lambda\intdx{  V\left(\frac{\delta}{\delta J(y)}\right)}{y}} \left[\frac{\delta}{\delta J(x)} K_k[J]\right] e^{\lambda\intdx{  V\left(\frac{\delta}{\delta J(y)}\right)}{y}}V'\left(\frac{\delta}{\delta J(x)}\right) }{x}\,.
\end{eqnarray}
We need then to evaluate the second operator in this expression. Let us iterate and introduce the operator
\be
\mathcal O_{x}[ \mu] = e^{-\mu\lambda\intdx{  V\left(\frac{\delta}{\delta J(y)}\right)}{y}} \left[\frac{\delta}{\delta J(x)} K_k[J]\right] e^{\mu\lambda\intdx{  V\left(\frac{\delta}{\delta J(y)}\right)}{y}}\,,
\ee
such that
\be
\frac{\dd\mathcal O_{x}}{\dd \mu}[ \mu] = -\lambda\  e^{-\mu\lambda\intdx{  V\left(\frac{\delta}{\delta J(y)}\right)}{y}} \left[\intdx{ V\left(\frac{\delta}{\delta J(y)}\right)}{y},\frac{\delta}{\delta J(x)} K_k[J]\right] e^{\mu\lambda\intdx{  V\left(\frac{\delta}{\delta J(y)}\right)}{y}}\,.
\ee
Now repeating the same game we have:
\begin{eqnarray}
&&-\left[\sum_{n=3}^\infty a_n \left(\frac{\delta}{\delta J(y)}\right)^{n}, \left(\frac{\delta}{\delta J(x)} K_k[J]\right)\right]\varphi[J]\,,\nonumber\\
&=&- \sum_{n=3}^\infty a_n \left[\left(\frac{\delta}{\delta J(y)}\right)^{n} \left(\frac{\delta}{\delta J(x)} K_k[J]\right)\varphi[J]- \left(\frac{\delta}{\delta J(x)} K_k[J]\right)\left(\frac{\delta}{\delta J(y)}\right)^{n}\varphi[J]\right]\,,\nonumber\\ 
&=&-\sum_{n=3}^\infty a_n \left[\left(\sum_{j=0}^n\begin{pmatrix}n\\j\end{pmatrix}\left(\frac{\delta}{\delta J(y)}\right)^{j} \left(\frac{\delta}{\delta J(x)} K_k[J]\right)\left(\frac{\delta}{\delta J(y)}\right)^{n-j}\varphi[J]\right)- \left(\frac{\delta}{\delta J(x)} K_k[J]\right)\left(-\frac{\delta}{\delta J(y)}\right)^{n}\varphi[J]\right]\,,\nonumber\\
&=&-\sum_{n=3}^\infty a_n \left[\sum_{j=1}^n\begin{pmatrix}n\\j\end{pmatrix}\left(\frac{\delta}{\delta J(y)}\right)^{j} \left(\frac{\delta}{\delta J(x)} K_k[J]\right)\left(\frac{\delta}{\delta J(y)}\right)^{n-j}\varphi[J]\right]\,,\nonumber\\
&=&-\sum_{n=3}^\infty a_n \left[n\frac{\delta}{\delta J(y)} \left(\frac{\delta}{\delta J(x)} K_k[J]\right)\left(\frac{\delta}{\delta J(y)}\right)^{n-1}\varphi[J]\right]\,,\nonumber\\
&=& -\frac{\delta}{\delta J(y)} \left(\frac{\delta}{\delta J(x)} K_k[J]\right)V'\left(\frac{\delta}{\delta J(y)}\right)\varphi[J]\,.
\end{eqnarray}
Using Eq.~\eqref{eq:xy2nd} we then obtain
\begin{eqnarray}
-\left[\intdx{ V\left(\frac{\delta}{\delta J(y)}\right)}{y},\frac{\delta}{\delta J(x)} K_k[J]\right] &=& -\int\dd^4 y \left(\frac{\delta}{\delta J(y)}\frac{\delta}{\delta J(x)} K_k[J]\right)V'\left(\frac{\delta}{\delta J(y)}\right)\,,
\end{eqnarray}
and
\begin{eqnarray}
\frac{\dd\mathcal O_{x}}{\dd \mu}[ \mu] &=& -\lambda\int\dd^4 y\left(\frac{\delta}{\delta J(y)}\frac{\delta}{\delta J(x)} K_k[J]\right)V'\left(\frac{\delta}{\delta J(y)}\right)\,. 
\end{eqnarray}
This leads to
\begin{eqnarray}
\mathcal O_{x}[1]  &=& \frac{\dd\mathcal O_{x}}{\dd \mu} + \mathcal O_{x}[0]\,,\nonumber\\
&=& -\lambda\intdx{\left(\frac{\delta}{\delta J(y)}\frac{\delta}{\delta J(x)} K_k[J]\right)V'\left(\frac{\delta}{\delta J(y)}\right)}{y}+\frac{\delta}{\delta J(x)} K_k[J]\,.
\end{eqnarray}
Inserting this in Eq.~\eqref{eq:dOdlambda} we get
\begin{eqnarray}\label{eq:dOdlambdanew}
\frac{\dd\mathcal O}{\dd\lambda} &=& -\intdx{ \left(\frac{\delta}{\delta J(x)}\right)^2 K_k[J] V''\left(\frac{\delta}{\delta J(x)}\right)}{x}\nonumber-\intdx{ \mathcal O_{x}[1]V'\left(\frac{\delta}{\delta J(x)}\right) }{x}\,,\nonumber\\
&=&-\intdx{ \left(\frac{\delta}{\delta J(x)}\right)^2 \frac{K_k[J]}{2} V''\left(\frac{\delta}{\delta J(x)}\right)}{x}\nonumber-\intdx{ \frac{\delta}{\delta J(x)} K_k[J]V'\left(\frac{\delta}{\delta J(x)}\right) }{x}\nonumber\\
&&+\lambda\intdx{\left(\frac{\delta}{\delta J(y)}\frac{\delta}{\delta J(x)} K_k[J]\right)V'\left(\frac{\delta}{\delta J(y)}\right)V'\left(\frac{\delta}{\delta J(x)}\right)}{x,y}\,.
\end{eqnarray}
We can now integrate Eq.~\eqref{eq:dOdlambdanew} between 0 and 1 to obtain
\begin{eqnarray}
\mathcal O[1] &=&-
\intdx{
\left(\frac{\delta}{\delta J(x)}\right)^2 \frac{K_k[J]}{2} V''\left(\frac{\delta}{\delta J(x)} \right)}{x}
\nonumber-\intdx{ \frac{\delta}{\delta J(x)} K_k[J]V'\left(\frac{\delta}{\delta J(x)}\right) }{x}\nonumber\\
&&+\frac{1}{2}\intdx{\left(\frac{\delta}{\delta J(y)}\frac{\delta}{\delta J(x)} K_k[J]\right)V'\left(\frac{\delta}{\delta J(y)}\right)V'\left(\frac{\delta}{\delta J(x)}\right)}{x,y} + K_k[J]\,.\nonumber\\
\end{eqnarray}
From Eq.~\eqref{eq:flow2}, using $\langle \hat \calO (\phi^k) \rangle \equiv \hat \calO (\delta/\delta J)\, \log Z^k[J]$, we finally obtain the required result which is used in the main body of the text:
\begin{eqnarray}
\partial_k \Gamma_k[\phi] &=& -  \frac{\mathcal O[1] Z^k[J]}{ Z^k[J]}\,,\nonumber\\
&=&
\intdx{
\left\langle \left(\frac{\delta}{\delta J(x)}\right)^2 K_k[J] V''\left(\phi^k(x)\right)\right\rangle\nonumber
}{x}
+\intdx{\left\langle \frac{\delta}{\delta J(x)} K_k[J]V'\left(\phi^k(x)\right) \right\rangle}{x}\nonumber\\
&&-\frac{1}{2}\intdx{\left\langle\left(\frac{\delta}{\delta J(y)}\frac{\delta}{\delta J(x)} K_k[J]\right)V'\left(\phi^k(x)\right)V'\left(\phi^k(y)\right)\right\rangle}{x,y} -K_k[J]\,.
\label{eq:appresult_orig}
\end{eqnarray}

\label{app:result}

\section{Flow Equation beyond the LPA}
Now, let us substitute in this equation the classical action by the coarse-grained effective action at scale $k$. To do so, we need to separate the free propagation part of the action from the interacting part by defining
\bea\label{eq:separation}
\Gamma[\phi]-\intdx{J\phi}{}=\Gamma[\bar\phi]+\intdx{ \left(\ddelta{\Gamma[\bar\phi]}{\bar\phi}}{}-J\right)\delta\phi+\frac{1}{2}\int\dd^4x  G^{-1}(\bar \phi;x,x)\delta\phi(x)^2 + \Gamma_{\rm I}[\delta\phi;\bar\phi]\,.
\eea

\be
\partial_k \Gamma^k[\phi] ~=~ - \partial_k \ln Z^k[J] = -  \frac{e^{-\intdx{  V\left(\frac{\delta}{\delta J}\right)}{}} K^k[J] e^{\intdx{  V\left(\frac{\delta}{\delta J}\right)}{}} Z^k[J]}{ Z^k[J]}\,.
\ee
\section{Adding Fermions}\label{app:fermions}

Considering a Lagrangian that includes fermions, one can write the partition function as
\be
Z[J_\phi,J_\eta,J_{\bar\eta}]~=~N\int\mathcal D\phi\mathcal D\eta\mathcal D\bar\eta\exp\left\{-
\intdx{
\frac{\partial\phi^2}{2}+\frac{m_\phi^2}{2}\phi^2 + \bar \eta \cancel\partial \eta + m_\eta \bar\eta\eta + \mathcal V_{\rm I}(\phi, \eta,\bar\eta)}{}
+\intdx{ J_\phi\phi}{}
+\intdx{ J_\eta\eta}{}
+\intdx{ \bar\eta J_{\bar\eta}}{}
\right\}~,
\ee
It is trivial to regulate the free partition function of such fermions by writing (in Fourier space)
\begin{eqnarray}
Z_0[\tilde{J_\eta}, {\tilde{J}}_{\bar\eta}] 
&~=~& Z_0[0] \exp\left\{\frac{1}{2}\int\frac{d^4 p}{(2\pi)^4}{\tilde{J}}_{\bar\eta}(p)(-i\cancel{p}+m_\eta)\Delta_F(p){\tilde{J}}_{\eta}(-p)\right\}\,,
\end{eqnarray}
where $\Delta_F(p)\equiv (p^2+m_\eta^2)^{-1}$ denotes the usual Feynman propagator. After regulating this propagator in the same manner as we did in Eq.~\eqref{eq:JJ}, we thus obtain
\be
Z_0^k[\tilde{J_\eta}, {\tilde{J}}_{\bar\eta}] ~=~ Z_0[0] \exp\left\{\frac{1}{2}\int \frac{\dd^4 p}{(2\pi)^4} \int_0^\infty \dd t\ \mathcal G^k(t) {\tilde{J}}_{\bar\eta}(-p)(-i\cancel{p}+m_\eta){\tilde{J}}_{\eta}(p) e^{-t(p^2+m_\eta^2)}\right\}\,.
\label{eq:JJfermions}\ee
From this definition, we can infer the expression of the associated $K^k[\tilde{J_\eta}, {\tilde{J}}_{\bar\eta}]$ function defined as $\partial_k Z_0^k[\tilde{J_\eta}, {\tilde{J}}_{\bar\eta}]\equiv K^k[\tilde{J_\eta}, {\tilde{J}}_{\bar\eta}]Z_0^k[\tilde{J_\eta}, {\tilde{J}}_{\bar\eta}]$:
\beq
K^k[\tilde{J_\eta}, {\tilde{J}}_{\bar\eta}]~=~\frac{1}{2}\int \frac{\dd^4 p}{(2\pi)^4} 
\int_0^\infty \dd t\ \partial_k \calG^k(t) {\tilde{J}}_{\bar\eta}(-p)(-i\cancel{p}+m_\eta){\tilde{J}}_{\eta}(p) e^{-t(p^2+m_\eta^2)} 
\,.
\eeq
As in the bosonic case, we use the expression of the full partition function to obtain the regularized partition function including interactions\footnote{Note the difference of sign in front of the derivative $\delta/\delta J_{\bar\eta}$ coming from the necessity for it to commute with $\bar\eta$ when acting on the free partition function.}
\be
Z^k[J_\phi,J_\eta,J_{\bar\eta}] = e^{-\intdx{\mathcal V_I\left(\frac{\delta}{\delta J_\phi},\frac{\delta}{\delta J_\eta},-\frac{\delta}{\delta J_{\bar\eta}}\right)}{}}Z_0^k[J_\phi]Z_0^k[J_\eta,J_{\bar\eta}]\,,
\ee
and after a very similar procedure as in the scalar case, we obtain
\begin{eqnarray}
\partial_k \Gamma_k[\phi,\eta,\bar\eta] &=&  \intdx{\left\langle\left(\frac{\delta}{\delta J_\phi(x)}\frac{\delta}{\delta J_{\phi}(x)}K_k[J_\phi]\right) \partial_\phi^2\mathcal V_{\rm I}\right\rangle}{x}-\intdx{\left\langle\left(\frac{\delta}{\delta J_\eta(x)}\frac{\delta}{\delta J_{\bar\eta}(x)}K_k[J_\eta, {J}_{\bar\eta}]\right) \partial_\eta\partial_{\bar\eta}\mathcal V_{\rm I}\right\rangle}{x} \nonumber\\
&+& \intdx{ \left\langle\frac{\delta}{\delta J_\phi(x)} K_k[J_\phi]\partial_{\phi}\mathcal V_{\rm I} \right\rangle}{x} +\intdx{\left\langle \frac{\delta}{\delta J_\eta(x)} K_k[J_\eta, {J}_{\bar\eta}]\partial_{\eta}\mathcal V_{\rm I}\right\rangle }{x} -\intdx{ \left\langle\frac{\delta}{\delta J_{\bar\eta}(x)} K_k[J_\eta, {J}_{\bar\eta}]\partial_{\bar\eta}\mathcal V_{\rm I}\right\rangle }{x} \nonumber\\
&-&\frac{1}{2}\intdx{\left\langle\left(\frac{\delta}{\delta J_\phi(x)}\frac{\delta}{\delta J_\phi(y)} K_k[J_\phi]\right)\partial_\phi\mathcal V_{\rm I}\left.{}\right|_{x}\partial_\phi\mathcal V_{\rm I}\left.{}\right|_{y}\right\rangle}{x,y}+\intdx{\left\langle\left(\frac{\delta}{\delta J_\eta(x)}\frac{\delta}{\delta J_{\bar\eta}(y)} K_k[J_\eta, {J}_{\bar\eta}]\right)\partial_\eta\mathcal V_{\rm I}\left.{}\right|_{x}\partial_{\bar\eta}\mathcal V_{\rm I}\left.{}\right|_{y}\right\rangle}{x,y}\nonumber\\
&+&K_k[J_\phi]+K_k[J_\eta, {J}_{\bar\eta}]\,.
\label{eq:appresultfermions}
\end{eqnarray}

Let us now do two things: $(i)$ Study the evolution of the effective action in a background of fields $(\phi_c,\eta_c,\bar\eta_c)$ solutions of the classical equation of motions, and $(ii)$ express the interacting part of the potential as a function of the full potential. We obtain
\begin{eqnarray}
\mathcal V_{\rm I} &=& V(\phi_c +\delta \phi,\eta_c+\delta\eta,\bar\eta_c+\delta\bar\eta) - \partial_{\phi} V(\phi_c \eta_c,{\bar\eta}_c)\delta\phi - \partial_{\eta} V(\phi_c \eta_c,\bar\eta_c)\delta\eta- \partial_{\bar\eta} V(\phi_c \eta_c,\bar\eta_c)\delta\bar\eta \nonumber\\
&-& \partial_\phi^2 V(\phi_c \eta_c,\bar\eta_c) \frac{{\delta\phi}^2}{2} - \partial_\eta\partial_{\bar\eta}V(\phi_c \eta_c,\bar\eta_c)\delta\eta\delta\bar\eta - \partial_\phi\partial_{\bar\eta}V(\phi_c \eta_c,\bar\eta_c)\delta\phi\delta\bar\eta- \partial_\phi\partial_{\eta}V(\phi_c \eta_c,\bar\eta_c)\delta\phi\delta\eta\,.
\end{eqnarray}
This leads to the following expressions for the derivatives
\begin{eqnarray}
\partial_{\delta\phi}\mathcal V_{\rm I} &=& \partial_\phi V(\phi_c +\delta \phi,\eta_c+\delta\eta,\bar\eta_c+\delta\bar\eta) - \partial_{\phi} V(\phi_c \eta_c,{\bar\eta}_c) - \partial_\phi^2 V(\phi_c \eta_c,\bar\eta_c){\delta\phi}  - \partial_\phi\partial_{\bar\eta}V(\phi_c \eta_c,\bar\eta_c)\delta\bar\eta- \partial_\phi\partial_{\eta}V(\phi_c \eta_c,\bar\eta_c)\delta\eta\nonumber\,,\\
\partial_{\delta\eta}\mathcal V_{\rm I} &=& \partial_\eta V(\phi_c +\delta \phi,\eta_c+\delta\eta,\bar\eta_c+\delta\bar\eta) - \partial_{\eta} V(\phi_c \eta_c,\bar\eta_c) - \partial_\eta\partial_{\bar\eta}V(\phi_c \eta_c,\bar\eta_c)\delta\bar\eta - \partial_\phi\partial_{\eta}V(\phi_c \eta_c,\bar\eta_c)\delta\phi\,,\nonumber\\
\partial_{\delta\bar\eta}\mathcal V_{\rm I} &=& \partial_{\bar\eta} V(\phi_c +\delta \phi,\eta_c+\delta\eta,\bar\eta_c+\delta\bar\eta) - \partial_{\bar\eta} V(\phi_c \eta_c,\bar\eta_c) - \partial_\eta\partial_{\bar\eta}V(\phi_c \eta_c,\bar\eta_c)\delta\eta - \partial_\phi\partial_{\bar\eta}V(\phi_c \eta_c,\bar\eta_c)\delta\phi\,,
\end{eqnarray}
and
\begin{eqnarray}
\partial_{\delta\phi}\partial_{\delta\phi}\mathcal V_{\rm I} &=& \partial_\phi^2 V(\phi_c +\delta \phi,\eta_c+\delta\eta,\bar\eta_c+\delta\bar\eta)  - \partial_\phi^2 V(\phi_c \eta_c,\bar\eta_c) \nonumber\,,\\
\partial_{\delta\bar\eta}\partial_{\delta\eta}\mathcal V_{\rm I} &=& \partial_{\bar\eta}\partial_{\eta}V(\phi_c +\delta \phi,\eta_c+\delta\eta,\bar\eta_c+\delta\bar\eta)  - \partial_\eta\partial_{\bar\eta}V(\phi_c \eta_c,\bar\eta_c)\,,
\nonumber\\
\partial_\phi\partial_{\delta\bar\eta}\mathcal V_{\rm I} &=& \partial_\phi\partial_{\bar\eta}V(\phi_c +\delta \phi,\eta_c+\delta\eta,\bar\eta_c+\delta\bar\eta)  - \partial_\phi\partial_{\bar\eta}V(\phi_c \eta_c,\bar\eta_c)\,. 
\end{eqnarray}
In full generality, these expressions are sufficient to obtain the flow equation of the effective action and all its derivatives with respect to each of the three fields. Our general results of Eq.~\eqref{eq:appresultfermions} can therefore be used by performing the shift
\be
(\phi,\eta,\bar\eta)\longrightarrow(\phi_c+\delta\phi,\eta_c+\delta\eta,\bar\eta_c+\delta\bar\eta)
\ee
and performing derivatives of $\Gamma^k$ with respect to $(\delta\phi,\delta\eta,\delta\bar\eta)$ before sending the latter to zero.
If one is interested in the flow equation for the scalar potential vacuum expectation value, this equation simplifies significantly as we can take the limit $(\eta_c,\bar\eta_c)\to (0,0)$, which gives 
\begin{eqnarray}
\partial_k \left[\partial_\phi {U^k}\left(\bar \phi, 0, 0\right)\right] &~=~& \frac{\partial_\phi^3 {U^k}(\bar \phi, 0, 0)}{32\pi^2}\int_0^\infty \frac{\dd t}{t^2}\ \mathcal \partial_k \calG^k(t)\,  e^{-t\ \partial_\phi^2 {U^k}(\bar \phi, 0, 0)} \nonumber\\
&-& \frac{\partial_\phi\partial_{\eta}\partial_{\bar\eta} {U^k}(\bar \phi, 0, 0)}{32\pi^2}{\partial_{\eta}\partial_{\bar\eta} {U^k}(\bar \phi, 0, 0)}\,\int_0^\infty \frac{\dd t}{t^2}\ \mathcal \partial_k \calG^k(t)\,  e^{-t\ \left[\partial_{\eta}\partial_{\bar\eta} {U^k}(\bar \phi, 0, 0)\right]^2}\,.
\label{eq:RGeq_fermions}
\end{eqnarray}

\bibliographystyle{apsrev4-2}

\bibliography{draft.bib}
\end{document}